\documentclass[twocolumn,superscriptaddress]{revtex4} 
\pdfoutput=1
\usepackage{graphicx}
\usepackage{color}

\usepackage{hyperref}
\usepackage{amsmath}

\usepackage{amssymb}
\usepackage{float}
\usepackage{natbib}
\usepackage[normalem]{ulem}

\begin{document}
\title{Coupled M{\"o}bius Maps as a Tool to Model Kuramoto Phase Synchronization}
\author{Chen Chris Gong}
\email[]{cgong@uni-potsdam.de}
\affiliation{Institute of Physics and Astronomy, University of Potsdam, Karl-Liebknecht-Stra\ss e 32, 14476 Potsdam, Germany}

\author{Ralf Toenjes}
\email[]{ralf.toenjes@uni-potsdam.de}
\affiliation{Institute of Physics and Astronomy, University of Potsdam, Karl-Liebknecht-Stra\ss e 32, 14476 Potsdam, Germany}

\author{Arkady Pikovsky}
\email[]{pikovsky@uni-potsdam.de}
\affiliation{Institute of Physics and Astronomy, University of Potsdam, Karl-Liebknecht-Stra\ss e 32, 14476 Potsdam, Germany}
\affiliation{Department of Control Theory, Nizhny Novgorod State University,
Gagarin Avenue 23, 606950 Nizhny Novgorod, Russia}

\date{\today}

\begin{abstract}
We propose M{\"o}bius maps as a tool to model synchronization phenomena in coupled phase oscillators.
Not only does the map provide fast computation of phase synchronization, it also reflects the underlying group
structure of the sinusoidally coupled continuous phase dynamics. 
We study map versions of various known continuous-time collective dynamics, such as 
the synchronization transition in the Kuramoto-Sakaguchi model of non-identical oscillators,
chimeras in two coupled populations of identical phase oscillators, and Kuramoto-Battogtokh chimeras on a ring, 
and demonstrate similarities and differences between the iterated map models and 
their known continuous-time counterparts.
\end{abstract}

\maketitle

\begin{quotation}
\end{quotation}

\section{Introduction} \label{intro}
Ensembles of sinusoidally coupled phase oscillators \cite{Kuramoto75, Pazo14}
are widely adopted as canonical models for synchronization in various scientific 
and engineering inquiries.
For instance, models of coupled phase oscillators have been 
successfully applied to functional connectivity of the human brain \cite{Cabral11, Petkoski18}, 
neuronal oscillatory behavior \cite{Daffertshofer10, Pazo18},
and neural encoding \cite{Doesburg09, Malagarriga15, Soman18}. 
Increasingly, they also serve as a computational tool in machine learning and 
artificial intelligence based on oscillatory neural networks \cite{Hoppensteadt00, Chakraborty14, Vodenicarevic16, Zhang19},
which opens up a new perspective for hardware implementations~\cite{Heger-Krischer-16}.

The most popular models in the field, the Kuramoto-Sakaguchi model \cite{Sakaguchi86} 
and the Winfree model \cite{Winfree} are formulated as systems
of ordinary differential equations for coupled phase oscillators. 
The goal of this paper is to formulate a discrete-time analogue of the Kuramoto-Sakaguchi model as a system
of coupled maps, that has similar dynamical properties but provides a fast computation of the dynamics in discrete-time steps. 

Globally coupled maps 
\cite{Kaneko-90,KANEKO91,Nozawa-92,Pikovsky-Kurths-94a,Just95,Topaj-Kye-Pikovsky-01} have been 
intensively studied in the literature, often with emphasis on the
collective dynamics of intrinsically chaotic units. Among the existing map models, 
coupled circle maps are ideal for studying synchronization phenomena due to their periodic domains. 
The simplest and most widely used circle map is the sine circle map
$\varphi \to \varphi+\Omega +\varepsilon\sin\varphi$, which has been explored in the context of 
global coupling \cite{KANEKO91, Chatterjee96, Osipov02} as well as in non-trivial coupling networks 
such as computational neural networks \cite{Bauer09}. However, the coupled sine circle maps have several properties different from that
of the Kuramoto-Sakaguchi model. For example, a known property of the continuous-time Kuramoto-Sakaguchi model 
is that clustering, i.e., the formation of several distinct synchronized groups, 
cannot occur \cite{Gong19}. However, for the sine circle maps, even though the map parameters can be 
tuned to certain regions such that no chaos is produced by the iteration of a single map (i.e., the mapping remains one-to-one), 
the coupled iterated map dynamics of identical units governed by the same mean field nevertheless produces various complex cluster states. 

As shown in previous literature, the propagator of continuous-time phase oscillators forced proportionally to the first harmonics of the phase
has the form of a M{\"o}bius 
transformation \cite{Marvel-Mirollo-Strogatz-09}. The M{\"o}bius transform lies at the heart of the low-dimensional dynamical 
theory for globally forced populations of continuous-time phase oscillators formulated by Watanabe and Strogatz 
(WS)~\cite{Watanabe94, Pikovsky08, Marvel-Mirollo-Strogatz-09, Pikovsky15, Chen17}. There, the M{\"o}bius 
transform is used to convert the original phase variables to new conserved quantities, 
such that the time-varying transformation parameters obey a simple low-dimensional system of ordinary differential equations.

In this paper, we implement a M{\"o}bius map, inspired from the aforementioned M{\"o}bius transform, as the basic circle map. 
The main arguments for studying synchronization and collective 
dynamics using M{\"o}bius maps are threefold. First, 
similar to the solution of a continuous-time Kuramoto-type phase model, 
an ensemble of infinite units governed by M{\"o}bius maps possesses 
a low-dimensional 
manifold (corresponding to the Ott-Antonsen (OA) manifold~\cite{Ott08} for continuous-time
oscillators). Therefore, the equation for the mean field
can be reduced to a low-dimensional map, whereas the mean-field equations 
for more general 
circle maps are generic infinite-dimensional 
nonlinear Perron-Frobenius operators. One exception is 
homographic maps~\cite{Griniasty-Hakim-94},
which as we will discuss in Section \ref{sec:rvr} below, are equivalent to M\"obius maps.
Secondly, iterated maps allow for large changes of the system state at each time step, 
in contrast to numerical 
integration of ordinary differential equations (ODEs) -- a property which can be exploited to speed up computation for 
large systems. Third, while M\"obius maps can fully reproduce the ODE 
behaviour within certain limit of map parameters, new and interesting dynamics is also possible, e.g. for strong negative coupling.

The plan of the paper is as follows. In Sec.~\ref{sec:map}, we first review the general form of the complex M{\"o}bius map and discuss its group properties. We discuss its single-map dynamics under
function iteration and fixed parameters. Next, by allowing the parameters of the map to vary in time and applying the group properties, we study the low-dimensional dynamics in globally coupled identical maps, and make the connection to the WS and OA mean-field reduction theories. Finally, we give a real-valued representation of the M\"obius map on the unit circle to be used in numerical calculations and remark on a connection of the theory of M\"obius maps with earlier results for homographic maps. In Sec.~\ref{sec:AdlerandKura} we give a compact expression of the M{\"o}bius map which solves the 
Adler equation for phase dynamics on an arbitrary time interval.  In Sec.~\ref{sec:mfd}, we
use this result to construct a map model of globally coupled, non-identical oscillators, as a discrete-time counterpart to the Kuramoto-Sakaguchi model~\cite{Sakaguchi86}.  
We discuss the dynamics of globally coupled M\"obius maps with frequency heterogeneity, of chimera states in two populations of identical 
phase oscillators with different intra- and inter-population couplings \cite{Abrams08}, and of chimeras on a periodic lattice of identical oscillators with non-local coupling 
\cite{ChimeraRing, Kura_chimera, Kalle17}. In all examples, known behaviours of the continuous-time dynamics can be reproduced qualitatively under positive coupling, and an interesting new synchronizing behaviour 
can be found for finite negative coupling, under which the familiar continuous-time dynamics would simply be incoherent or asynchronous.

\section{M\"obius map and its properties}
\label{sec:map}
In this section we heavily rely on the excellent introduction
of M\"obius transformation and M\"obius group in the context of
continuous-time dynamics
by Marvel, Mirollo, and Strogatz~\cite{Marvel-Mirollo-Strogatz-09}. We will repeat some results from Ref.~\cite{Marvel-Mirollo-Strogatz-09} for the sake of consistency. Our extension
of Ref.~\cite{Marvel-Mirollo-Strogatz-09} is the introduction of the M\"obius circle map as
a dynamical system (Section~\ref{sec:sim}), and the formulation of the low-dimensional
discrete-time equations governing the evolution of ensembles (Section~\ref{sec:lowDMM}).
Additionally, we give a real-valued representation of the map on the complex unit circle and connect the M\"obius maps to homographic maps on the real line, where 
low-dimensional mean-field behavior on an invariant manifold for infinite ensembles has been reported previously~\cite{Griniasty-Hakim-94}.

\subsection{Standard form and dynamics of a single iterated map}
\label{sec:sim}

For the M{\"o}bius transformation we use the same parametrization as in Ref.~\cite{Marvel-Mirollo-Strogatz-09}
\begin{equation}
	\mathcal{M}_{q,\psi}\left(z\right) = \frac{q + e^{i\psi}z}{1 + q^*e^{i\psi}z}~,
	\label{eq:mm}
\end{equation}
with parameters $(q,\psi) \in \mathbb{D}\times S^1$, where $|q|<1$, i.e., $q$ is on the open complex unit disc $\mathbb{D}$, and $\exp(i\psi)\in S^1$ on the complex unit circle $S^1$. The complex conjugate of $q$ is denoted as $q^*$. Transformation~\eqref{eq:mm} 
can be applied to any complex number $z$ in the closed unit disk, $z \in \{\mathbb{D} \cup S^1\}$. Transformation~\eqref{eq:mm} is invertible and leaves the complex unit circle invariant. This property
will be used to define a circle map.
M\"obius transformation \eqref{eq:mm} can be decomposed as two independent actions
\begin{equation}
	\mathcal{M}_{q,\psi}\left(z\right) = \mathcal{C}_q\circ\mathcal{R}_\psi\left(z\right)~,
\end{equation} 
where $\mathcal{R}_\psi$ denotes rotation by an angle $\psi$
\begin{equation}
\mathcal{R}_\psi: z \to e^{i\psi}z ~,
\end{equation}
and $\mathcal{C}_{q}$ denotes a directional contraction
\begin{equation}
	\mathcal{C}_{q}: z \to \frac{q+z}{1+q^*z}~.
\end{equation}
The identity transformation is $\mathcal{M}_{0,0}$. 
The rotational actions commute: $\mathcal{R}_{\psi_1}\circ \mathcal{R}_{\psi_2} =\mathcal{R}_{\psi_2}\circ \mathcal{R}_{\psi_1} 
= \mathcal{R}_{\psi_1+\psi_2}$, with the inverse of the rotation $\mathcal{R}^{-1}_\psi = \mathcal{R}_{-\psi}$. 
The inverse of the contraction is $\mathcal{C}_q^{-1} = \mathcal{C}_{-q}$ such that
\begin{equation}
	\mathcal{M}^{-1}_{q,\psi} = \mathcal{R}_{-\psi}\circ\mathcal{C}_{-q}~. 
\end{equation}
Rotational symmetry is expressed as
\begin{equation}
	\mathcal{C}_q = \mathcal{R}_{-\psi}\circ\mathcal{C}_{qe^{i\psi}} \circ\mathcal{R}_{\psi} ~.
\end{equation}

It is well known that M\"obius transformation forms a group under functional composition 
\begin{equation}
\mathcal{M}_{q_2,\psi_2}\left(z\right) = \mathcal{M}_{q_1,\psi_1}\circ \mathcal{M}_{q_0,\psi_0}\left(z\right)~.
\label{eq:gp}
\end{equation}
This is to say, the group parameters $q_2,\psi_2$ of the new transformation 
are functions of the parameters of the component transformations. 
Under parametrization \eqref{eq:mm}, these functions are (see Appendix~\ref{sec:mgp} for a detailed derivation): 
\begin{equation} \label{Eq:GroupParmTrans}
	q_2 = \mathcal{M}_{q_1,\psi_1}\left(q_0\right) \;,\quad
	e^{i\psi_2}  = \mathcal{C}_{q_1q^*_0}\left(e^{i\psi_1}\right) \cdot e^{i\psi_0}\;.
\end{equation}

We now consider a complex variable $z=e^{i\varphi}$ on the unit circle, and
define a M\"obius circle map by virtue of the transformation~\eqref{eq:mm}:
\begin{equation}\label{eq:mmds}
e^{i\varphi^{(n+1)}}=\mathcal{M}_{q,\psi}\left(e^{i\varphi^{(n)}}\right)\;,
\end{equation}
where $n=0,1,\ldots$ is a discrete time index. Map $\mathcal{M}_{q,\psi}$ is defined by 
a complex parameter $q=\rho\exp(i\vartheta)$ and a real parameter $\psi$.
Here we first assume the parameters $q, \psi$ to be constant.

\begin{figure}[t]
\centering\includegraphics[width=\columnwidth]{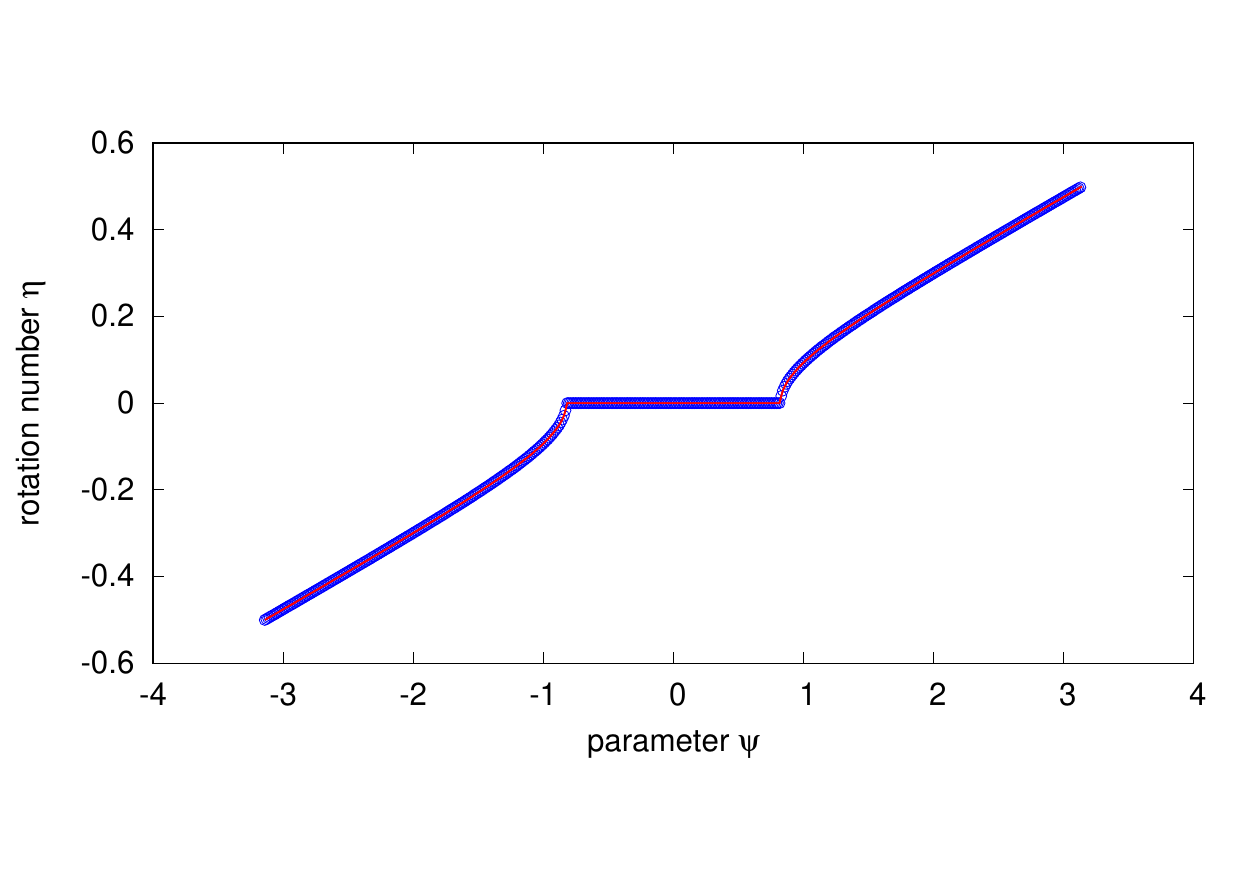}
\caption{The rotation number $\eta$ of iterated M\"obius maps as a function of map parameter $\psi$, given $\rho = 0.4$.
Open circles: direct numerical simulations; solid line: formula~\eqref{eq:rot_number2}.}
\label{fig:rn}
\end{figure}

Circle map \eqref{eq:mmds}, contrary to generic smooth circle maps, 
possesses a very simple dynamics under iteration. It can be shown that it has just one Arnold tongue 
-- a region of synchronous phase locking with rotation number zero, while tongues with other
rotation numbers do not exist. This should be contrasted with the existence
of Arnold tongues with all rational rotation numbers for generic circle maps~\cite{Katok-Hasselblatt-95}.
The rotation number $\eta$ can be defined according to Refs.~\cite{Katok-Hasselblatt-95,syncbook} as
\begin{equation}
\eta = \lim_{n\to\infty}\frac{\varphi^{(n)}-\varphi^{(0)}}{ 2\pi n}~,
\end{equation}
where the phase variable is lifted to the real line. Derived in detail in Appendix~\ref{sec:app:rn}, 
the rotation number of map \eqref{eq:mmds} can be expressed as 
\begin{equation} \label{eq:rot_number2}
\eta = \begin{cases}
    0&\text{for}\quad \rho > \left|\sin\frac{\psi}{2}\right|\;,\\
    \frac{1}{\pi}\arctan\left(\tan\frac{\psi}{2}\cdot\sqrt{1-\frac{\rho^2}{\sin^2\frac{\psi}{2}}}\right)&\quad
    \textrm{otherwise.}
\end{cases}
\end{equation}
The rotation number of the M\"obius map as a function of parameter $\psi$
is demonstrated in Fig.~\ref{fig:rn}, where numerical and analytical results are shown to 
coincide. One can see a plateau with $\eta = 0$ for 
$\rho > \left|\sin\frac{\psi}{2}\right|$ and smooth dependence on $\psi$
outside the plateau. In the domain $\eta \neq 0$, as is derived in Appendix~\ref{sec:app:rn},
the M\"obius map is conjugate (by virtue of another M\"obius transformation) to a circle shift.

The above calculation consists an analytical proof for the existence of one single Arnold tongue in iterated dynamics of M\"obius maps of fixed parameters. Another way of proving the same result is by considering the group property of M\"obius map~\eqref{eq:gp}. If we assume to the contrary, that M\"obius map \eqref{eq:mmds} under iteration has an Arnold tongue with a non-integer rotation number, then it follows that there exists a stable periodic orbit with a period larger than one. This in turn implies that an iteration over such a period results in several stable fixed points. But from the group property~\eqref{eq:gp}, we know that any iteration of the M\"obius map is again a M\"obius map, and a M\"obius map can have at most only one stable fixed point. This contradicts the assumption that M\"obius  map \eqref{eq:mmds} has an Arnold tongue with a non-integer rotation number. Hence M\"obius map \eqref{eq:mmds} can only possess Arnold tongues of integer rotation number. 

This special property of M\"obius maps having only one Arnold tongue has another
consequence, namely, that different phases iterated by the same M\"obius map may form at most
one cluster. A cluster $K\{\varphi^{(n)}_j, j=1\dots N\}$ is a subset of phases in an ensemble which contract to a single point on the unit circle in the course of the phase evolution, i.e., $\lim_{n\to\infty}\exp[i(\varphi_k^{(n)}-\varphi_j^{(n)})]=1$ for all pairs of phases $\{\varphi_k, \varphi_j\}$ in the cluster. Since the evolution of any phase
variable $\varphi^{(0)} \to \varphi^{(n)}$ is given by the same M\"obius map, which has at most one attractive fixed point, only one cluster can be formed
under a common iterated M\"obius map of fixed or time-varying parameters, 
with the possible exception of one phase located precisely at the unstable fixed point of the map, outside the cluster.

Under the iteration of M\"obius maps, as with all invertible circle maps, 
chaotic dynamics of the phases cannot occur, regardless of 
whether the map parameters are constant, time-varying, or follow chaotic sequences.

\subsection{Low-dimensional evolution of oscillator ensembles under M\"obius maps}
\label{sec:lowDMM}
The group property of the M\"obius maps, as shown by Eqs. \eqref{eq:gp} and \eqref{Eq:GroupParmTrans}, implies that the evolution under M\"obius map 
dynamics~\eqref{eq:mmds} from any set of initial states is reducible to a three-dimensional
evolution of the map parameters $q$ and $\psi$. To see this, consider 
the
single-map dynamics \eqref{eq:mmds} with 
an arbitrary discrete sequence of parameters $q^{(n)},\psi^{(n)}$ that vary in time
\begin{equation}\label{eq:indMM}
e^{i\varphi^{(n)}}=\mathcal{M}_{q^{(n)},\psi^{(n)}} \left(e^{i\varphi^{(n-1)}}\right)\;.
\end{equation}
Due to the group property \eqref{eq:gp}, the evolution over any time interval from the initial state $\exp(i\varphi^{(0)})$ to the final state $\exp(i\varphi^{(n)})$
can also be expressed as a M\"obius map
\begin{equation}\label{eq:fromIC}
	e^{i\varphi^{(n)}}=\mathcal{M}_{Q^{(n)},\Psi^{(n)}} \left(e^{i\varphi^{(0)}}\right).
\end{equation}
Shifting $n\to (n-1)$, applying map \eqref{eq:indMM} of time step $n$ to both sides of \eqref{eq:fromIC} and using 
the transformation rule \eqref{Eq:GroupParmTrans} for composite group parameters, we obtain the evolution equations of $Q$ and $\Psi$
\begin{equation} \label{eq:wsa}
\begin{aligned}
 Q^{(n)}&=\mathcal{M}_{q^{(n)},\psi^{(n)}} \left(Q^{(n-1)}\right) ~,\\
 e^{i\Psi^{(n)}}&=\mathcal{C}_{q^{(n)}Q^{*(n-1)}}\left(e^{i\psi^{(n)}}\right) e^{i\Psi^{(n-1)}}~.
\end{aligned}
\end{equation}
The initial values $Q^{(0)}=0$ and $\Psi^{(0)}=0$ follow from the identity map $\mathcal{M}_{0,0}(z)=z$. 

We note that the transformation governing $Q$ is the same as the original M\"obius
map~\eqref{eq:indMM} for the phase $\varphi$, but here it is applied 
to a complex variable defined on the unit disc, and not on the unit circle. 

Because
transformation~\eqref{eq:wsa} does not depend on the initial phase $\varphi^{(0)}$,
it can be used to describe the evolution of any initial state of the phase by
first applying the map~\eqref{eq:wsa} for $Q^{(n)}$ and then applying the map~\eqref{eq:fromIC} for the phase.
Thus, the evolution of any ensemble of oscillators 
by a sequence of M{\"o}bius maps is always restricted to a three-dimensional 
manifold described by \eqref{eq:wsa} and parametrized by $(Q^{(n)},\Psi^{(n)})\in \mathbb{D}\times S^1$. %
Hence, the discrete-time dynamics \eqref{eq:wsa} for an ensemble of oscillators under common forcing is fully analogous to the Watanabe-Strogatz quasi-mean-field 
equations in the continuous-time case \cite{Watanabe94}. 
The role of the three-dimensional manifold is the same as in the continuous-time case: It implies that for any number of units, the dynamics can be split into
constants of motion (e.g., initial values of the phases) and three dynamical variables $(Q^{(n)},\Psi^{(n)})$, the evolution of which may be nontrivial. One often calls this property of the phase dynamics ``partial integrability''.

Even for an infinite ensemble of oscillators described by a density $\mu(\varphi)$, the evolution takes place on a three-dimensional invariant manifold.
$\mu_{Q,\Psi}(\varphi)$ denotes the density of oscillator phases after the transformation $\exp(i\varphi) \to \mathcal{M}_{Q,\Psi}(\exp(i\varphi))$ from an initial density $\mu_{0,0}(\varphi)$. 
The family of densities $\mu_{Q,\Psi}(\varphi)$ is a three-dimensional invariant manifold,
parametrized by $Q$ and $\Psi$. In the special case of a continuous, 
uniform phase density $\mu_{0,0}(\varphi) = 1/(2\pi)$, the resulting family of densities $\mu_{Q,\Psi}(\varphi)=\mu_{Q}(\varphi)$ is independent of $\Psi$, which corresponds to an angular shift that leaves the uniform density invariant on the circle. 
As shown in Ref.~\cite{Marvel-Mirollo-Strogatz-09} and Appendix \ref{sec:OAemm}, $\mu_{Q}(\varphi)$ is the family of wrapped Cauchy distributions
\begin{equation}
\mu_Q(\varphi)=\frac{1}{2\pi}\frac{1-|Q|^2}{|e^{i\varphi}-Q|^2}\;.
\end{equation}
This family of phase densities is called the Ott-Antonsen (OA) manifold in continuous-time phase dynamics \cite{Ott08}, and we shall use the same name to discuss the manifold in the map dynamics here.
On the OA manifold, the Kuramoto mean field $Z = \left\langle \exp(i\varphi)\right\rangle_{\mu_Q}$ is in fact identical to the M\"obius map parameter $Q$ (Appendix \ref{sec:OAemm}).
Replacing map parameter $Q$ with the mean field $Z$ in \eqref{eq:wsa}, it follows that the exact evolution of the ensemble mean field can also be expressed as an iterated M\"obius transformation
\begin{equation} 
Z^{(n)}=\mathcal{M}_{q^{(n)},\psi^{(n)}}\left(Z^{(n-1)}\right)= \frac{q^{(n)}+e^{i\psi^{(n)}} Z^{(n-1)}}{1+q^{*(n)} e^{i\psi^{(n)}} Z^{(n-1)}}\;.\quad
\label{eq:oaeq}
\end{equation}
It is interesting to note that, the map for the mean field \eqref{eq:oaeq} has exactly the same form as the map describing the dynamics \eqref{eq:indMM} of every single oscillator in the ensemble. However, the domains of the two maps differ: while the complex oscillators $\exp(i\varphi)$ are always on the unit circle, the mean field $Z$ is an element of the unit disc, i.e. $|Z|\leq 1$.

As a side note, the sequences $q^{(n)}$ and $\psi^{(n)}$ in Eq.~\eqref{eq:indMM} are arbitrary and may be functions of $Z$ or contain random components. In this way, more complicated noisy dynamics and globally coupled oscillators can also be studied with the discrete map model proposed here for which all the results above still hold.

\subsection{Real-valued representation and relation to homographic maps}
\label{sec:rvr}

For computational purposes, an equivalent form of parametrization 
of the M\"obius map may be used, which is more suited
for programming languages that do not natively support a data type for complex numbers. Using identity 
$\exp(is)=(1+i\tan\frac{s}{2})(1-i\tan\frac{s}{2})^{-1}$ and $q=\rho\exp(i\vartheta)$ as before, we can rewrite~\eqref{eq:mmds} in the trigonometric form
\begin{equation}
\tan\frac{\varphi^{(n+1)}-\vartheta}{2}=\frac{1-\rho}{1+\rho}\tan\frac{\varphi^{(n)}+\psi-\vartheta}{2}~.
\label{eq:tan2}
\end{equation}
Manipulating \eqref{eq:tan2} we then obtain the computationally simple
M\"obius map via the $\text{ATAN2}$ function
\begin{equation}
\begin{gathered}
\varphi^{(n+1)}=\vartheta+\text{ATAN2}\left[(1-\rho^2)\sin(\varphi^{(n)}+\psi-\vartheta),\right.\\
\left.(1+\rho^2)\cos(\varphi^{(n)}+\psi-\vartheta)+2\rho\right]\;.
\end{gathered}
\label{eq:atan2}
\end{equation}

Griniasty and Hakim~\cite{Griniasty-Hakim-94} studied a family of homographic maps, defined for real variable $x$ as
\begin{equation}
x^{(n+1)}=a-\frac{b}{x^{(n)}}\;.
\label{eq:homm}
\end{equation}
This map leaves a Cauchy distribution invariant, in the same way that M\"obius maps leave a wrapped Cauchy distribution invariant. The homographic map \eqref{eq:homm} can be shown to be equivalent to a M\"obius map. Indeed,
by a substitution $x^{(n)}=\tan{(\varphi^{(n)}/2)}=i(1-e^{i\varphi^{(n)}})(1+e^{i\varphi^{(n)}})^{-1}$ we can rewrite \eqref{eq:homm} as
the M\"obius map~\eqref{eq:mmds} for $\varphi^{(n)}$ with parameters
\begin{equation}
q=\frac{i(1-b)+a}{i(1+b)+a},\quad 
e^{i\psi}=\frac{a-i(1+b)}{a+i(1+b)}\;.
\label{eq:frl}
\end{equation}
Given a Cauchy distribution with mean $r$ and scale parameter $s$, the complex number $z = r + is$ can be shown to be transformed under the same homographic map $z^{(n+1)}=a-b/z^{(n)}$ as the Cauchy distributed variables $x^{(n)}$. This is to say, that the low-dimensional reduction for ensembles evolved under 
iterated homographic maps was already deduced 
in \cite{Griniasty-Hakim-94}, almost at the same time as the 
discovery of the low-dimensional dynamics in the ensembles of phase oscillators \cite{Watanabe94,Ott08,Marvel-Mirollo-Strogatz-09}. 

\section{Relation to Adler equation}
\label{sec:AdlerandKura}
Many continuous-time phase models of coupled oscillators 
can be written in the form of an Adler equation \cite{Adler46} with constant or time-varying parameters. 
As we shall see, the solution of the Adler equation has the form of a M\"obius map.
This allows us to build M\" obius map  models of coupled oscillators analogous to continuous-time models. In this section we 
relate the parameters 
of the Adler equation to those of the M\"obius map.

The Adler equation with constant parameters can be written in the form
\begin{equation}
\dot\varphi=A\left[\lambda-\sin(\varphi-\beta)\right]~,
\label{eq:adleq}
\end{equation}
where the real-valued parameters consist of the amplitude $A$, the ratio $\lambda$ between the 
constant bias term and the amplitude of the sinusoidal forcing, and the phase shift $\beta$.
It is known that for $|\lambda|\leq 1$ the Adler equation has a steady state solution, and for $|\lambda|>1$, it yields phase rotations.

The solution of the Adler equation with fixed parameters $A, \lambda, \beta$ over a time interval $\tau $ can be shown  
to be a M\" obius map (see Appendix \ref{sec:km} for details). Here $A$ and $\tau$ only enter the solution as a product 
$P=A\tau$. Denoting 
\begin{equation}\label{eq:parm_map}
	\sigma = \sqrt{1-\lambda^2}, \quad \Gamma = \tanh\left(\frac{P}{2}\sigma\right) ~,
\end{equation}
and using the conventions $\sqrt{-1}=i$ and $\tanh(ix)=i\tan(x)$, we can write 
the solution of the Adler equation as
the M\"obius map 
\begin{equation} \label{eq:main_num_eqn}
	e^{i\varphi(\tau)} = \frac{(\sigma + i\lambda\Gamma)e^{i\varphi(0)} + e^{i\beta}\Gamma}{(\sigma - i\lambda\Gamma) + e^{i\varphi(0)} e^{-i\beta}\Gamma} = \mathcal{M}_{q,\psi}\left(e^{i\varphi(0)}\right) 
\end{equation}
with group parameters
\begin{equation}\label{eq:mmadl}
	q= e^{i\beta}\frac{\Gamma}{\sigma-i\lambda\Gamma}~, \qquad e^{i\psi} = \frac{\sigma+i\lambda\Gamma}{\sigma-i\lambda\Gamma}~. 
\end{equation}
The saddle-node bifurcation for the Adler equation at $|\lambda|=1$ corresponds to the tangent bifurcation of the circle map. At the bifurcation, 
Eqs.~\eqref{eq:mmadl} need to be evaluated in the limit $\lambda\to \pm 1$, i.e.,
\begin{equation}
    q = e^{i\beta} \frac{P}{2\mp i P}~, \qquad e^{i\psi} = \frac{2\pm i P}{2\mp i P}~.
\end{equation}
If the solution to the Adler equation after interval
$\tau$ is a M\"obius map, 
then the evolution under iterated M\"obius maps is a M\"obius map 
again, as shown by the group property 
in Sec.~\ref{sec:map}. Therefore, taking infinitesimal time steps, the solution of the Adler equation with time dependent parameters $A(t)$, $\lambda(t)$ and $\beta(t)$ is still a M\"obius map.

Consequently, all basic properties of the Adler equation are inherited by the 
M\"obius map. 
In particular, it is known that for a solution to the Adler equation which 
has a periodic dependence on its parameters, there is only one Arnold tongue
for every integer rotation number~\cite{Buchstaber2010,Ilyashenko2011}.
This matches exactly the property of M\"obius map as discussed in Section~\ref{sec:map}, 
i.e. the M\"obius map has at most one stable fixed point 
in the synchronized state.
Hence, Eq.~\eqref{eq:main_num_eqn} can be viewed as a numerical scheme to simulate the
continuous-time Adler equation with small time step $\tau$. In fact, while a standard Euler scheme, 
which to the linear order in $\tau$ coincides with the M\"obius map, breaks the Watanabe-Strogatz partial integrability
of the Adler equation~\cite{Gong19}, the M{\"o}bius map~\eqref{eq:main_num_eqn} preserves this partial integrability, similar to the symplectic integration schemes for Hamiltonian equations. Because the map keeps the properties of the Adler equation also for large $\tau$, it offers
a possibility to model features of oscillators obeying the Adler-type dynamics with an increased computational
efficiency, even though the main bottleneck of computing the mean-field at each step remains for coupled maps
(see Appendix \ref{sec:ce} for more details).

In the special case where only the amplitude $A=A(t)$ has explicit time dependence, the Adler equation \eqref{eq:adleq} has the form of a phase response to a time-dependent forcing, 
$\dot\varphi = H(\varphi)A(t)$. The exact solution of \eqref{eq:adleq} in this case can be obtained by separation of variables. The solution has the same form as \eqref{eq:parm_map} and \eqref{eq:main_num_eqn}, except in this case the parameter $P$ is the 
integral of $A(t)$ over the time interval $\tau$
\begin{equation}
	P = \int_0^\tau A(t)dt~.
\end{equation}
Here, the time-dependent kick amplitude $A(t)$ can be any generic function, e.g. a delta pulse or a constant force.

\section{Mean-Field Dynamics for Phases Evolved Under Coupled M\"obius Maps}
\label{sec:mfd}
\subsection{Formulation of a model of globally coupled maps with Kuramoto-Sakaguchi-type coupling} \label{sec:KS}
The Kuramoto-Sakaguchi model of  globally coupled phase oscillators
is formulated as a system of $N$ Adler-type equations
\begin{equation}\label{eq:kurasaka}
    \dot\varphi_j = \omega_j + \varepsilon R\sin(\Theta-\varphi_j-\alpha)~,
\end{equation}
where the forcing, common to all oscillators, is expressed through the complex mean field
\begin{equation}
    Z = Re^{i\Theta} = \frac{1}{N}\sum_{j=1}^{N} e^{i\varphi_j} = \langle e^{i\varphi_j}\rangle_j~.
\end{equation}
Here, natural frequencies $\omega_j$ can be typically assumed to be sampled from some distribution.
In this section, we build a discrete analogue of model \eqref{eq:kurasaka} based on the M\"obius maps.

Comparing \eqref{eq:kurasaka} with the Adler equation \eqref{eq:adleq}, one can see that $A$ corresponds
to $\varepsilon R$, $\beta$ corresponds to $\Theta-\alpha$, and $\lambda$ corresponds
to $\omega_j/(\varepsilon R)$. However, a direct application of the map solution \eqref{eq:main_num_eqn} of the 
discrete-time Adler equation is not optimal here, because parameters $\lambda$ and $A$ enter \eqref{eq:main_num_eqn} in a 
rather complex manner. In order to obtain a simple discrete-time model, which not only carries the essential properties of 
a globally forced continuous-time phase model, but also allows an analytic evaluation of averages with respect to some 
distribution of natural frequencies, we split the phase evolution in the Kuramoto-Sakaguchi
model into two stages. In the first stage a delta pulse of strength $P^{(n)}=\varepsilon R^{(n)}$ is applied to all oscillators,
the solution to which corresponds to \eqref{eq:main_num_eqn} with $\beta^{(n)}=\Theta^{(n)}-\alpha$ 
and $\lambda=0$. Here $R^{(n)}\exp(i\Theta^{(n)})$ is the mean field calculated just prior to the kick by the delta pulse.
In the second stage, the oscillators undergo free rotation for a time interval $T$ with individual natural frequencies $\omega_j$. 
This stage corresponds to the map $\varphi_j
 \to \varphi_j+\omega_j T$. Combining stages one and two, we formulate
 the resulting model of heterogeneous, globally coupled oscillators as 
\begin{equation}
e^{i\varphi_j^{(n+1)}}=e^{i\omega_j T}\frac{e^{i\varphi_j^{(n)}}+e^{i\Theta^{(n)}-i\alpha} \tanh\frac{\varepsilon R^{(n)}}{2}}
{1+ e^{i\varphi_j^{(n)}} e^{-i\Theta^{(n)}+i\alpha} \tanh\frac{\varepsilon R^{(n)}}{2} }\;.
\label{eq:gcmm}
\end{equation}
Here the mean field is defined as
\begin{equation}
   R^{(n)}e^{i\Theta^{(n)}}= \frac{1}{N}\sum_{j=1}^{N} e^{i\varphi_j^{(n)}}\;.
   \label{eq:mf}
\end{equation}
Taking $T\to dt$ and $\varepsilon\to\varepsilon dt$ the map \eqref{eq:gcmm} is to the linear order in $dt$ equivalent to an Euler integration step for the continuous-time Kuramoto-Sakaguchi model \eqref{eq:kurasaka} and solves the ODE exactly in the limit $dt\to 0$. In the thermodynamic limit and on the Ott-Antonsen manifold, the phase density 
$\mu_Q(\varphi,\omega)$ for each value of frequency $\omega$ is a wrapped Cauchy 
distribution with mean field 
$Q(\omega) = \left\langle e^{i\varphi} \right\rangle_{\mu_Q(\varphi,\omega)}$, as shown above in Sec.\,\ref{sec:lowDMM}.
According to \eqref{eq:oaeq}, parameter
$Q$ then obeys the same map as the individual phases having natural frequency $\omega$ \eqref{eq:gcmm}, i.e.
\begin{equation}\label{eq:kurasaka_mean_w}
	Q^{(n+1)}(\omega) = e^{i\omega T}\frac{Q^{(n)}(\omega)+e^{i(\Theta^{(n)}-\alpha)} \tanh\frac{\varepsilon R^{(n)}}{2}}
{1 + Q^{(n)}(\omega)e^{-i(\Theta^{(n)}-\alpha)}\tanh\frac{\varepsilon R^{(n)}}{2}}~.
\end{equation}
The value of the mean field $Z$ can be calculated as the average of $Q^{(n)}(\omega)$ with respect to a continuous distribution density of natural frequencies $g(\omega)$
\begin{equation}\label{eq:kurasaka_mean}
	Z^{(n)} = R^{(n)}e^{i\Theta^{(n)}} = \int_{-\infty}^\infty Q^{(n)}(\omega) g(\omega) d\omega~.
\end{equation}

Similar to the approach of Ott and Antonsen~\cite{Ott08}, we can assume that $Q(\omega)$ is analytic in the upper
half-plane, which allows us to calculate the integral via the residue theorem. For a Lorentzian 
frequency distribution of mean $\omega_0$ and scale parameter $\gamma$
\begin{equation}
	g(\omega) = \frac{1}{\pi\gamma} \frac{\gamma^2}{(\omega-\omega_0)^2+\gamma^2}~,
\end{equation}
the mean field is $Z=Q(\omega_0+i\gamma)$.
Accordingly, the global mean field evolves according to the following map
\begin{equation}\label{eq:KSMF}
	Z^{(n+1)} = e^{(i\omega_0-\gamma) T} \frac{Z^{(n)}+ e^{i(\Theta^{(n)}-\alpha)}\tanh\left(\frac{\varepsilon R^{(n)}}{2}\right)}{1 +  Z^{(n)} e^{-i(\Theta^{(n)}-\alpha)} \tanh\left(\frac{\varepsilon R^{(n)}}{2}\right)}~.
\end{equation}
The first part of map \eqref{eq:KSMF}, $e^{(i\omega_0-\gamma) T}$, consists of a rotation with the
mean frequency $\omega_0$ of the ensemble, and a decay of the mean field due to population heterogeneity 
$\gamma$, which is determined by the width of the natural frequency distribution. Equation~\eqref{eq:KSMF} is a discrete analogue of the Ott-Antonsen equation~\cite{Ott08}
which describes the dynamics of the mean field in the Kuramoto-Sakaguchi model. 

\begin{figure} [t]
\setlength{\unitlength}{1cm}
\begin{picture}(4.2,4.2)
\put(0,0){\includegraphics[height=4.5cm]{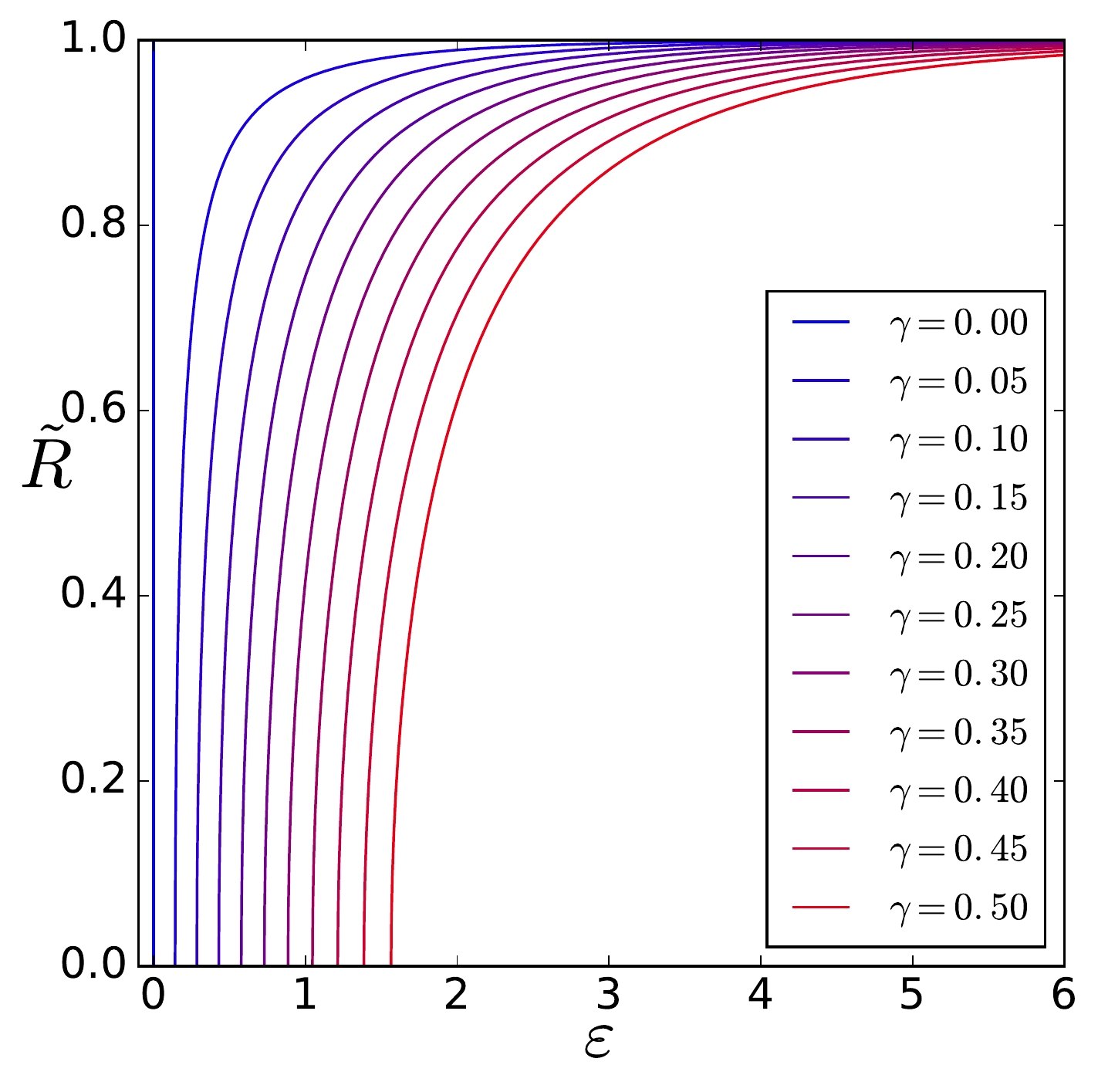}}
\put(0,3.8){\bf (a)}
\end{picture}
\begin{picture}(4.2,4.2)
\put(0.1,0){\includegraphics[height=4.5cm]{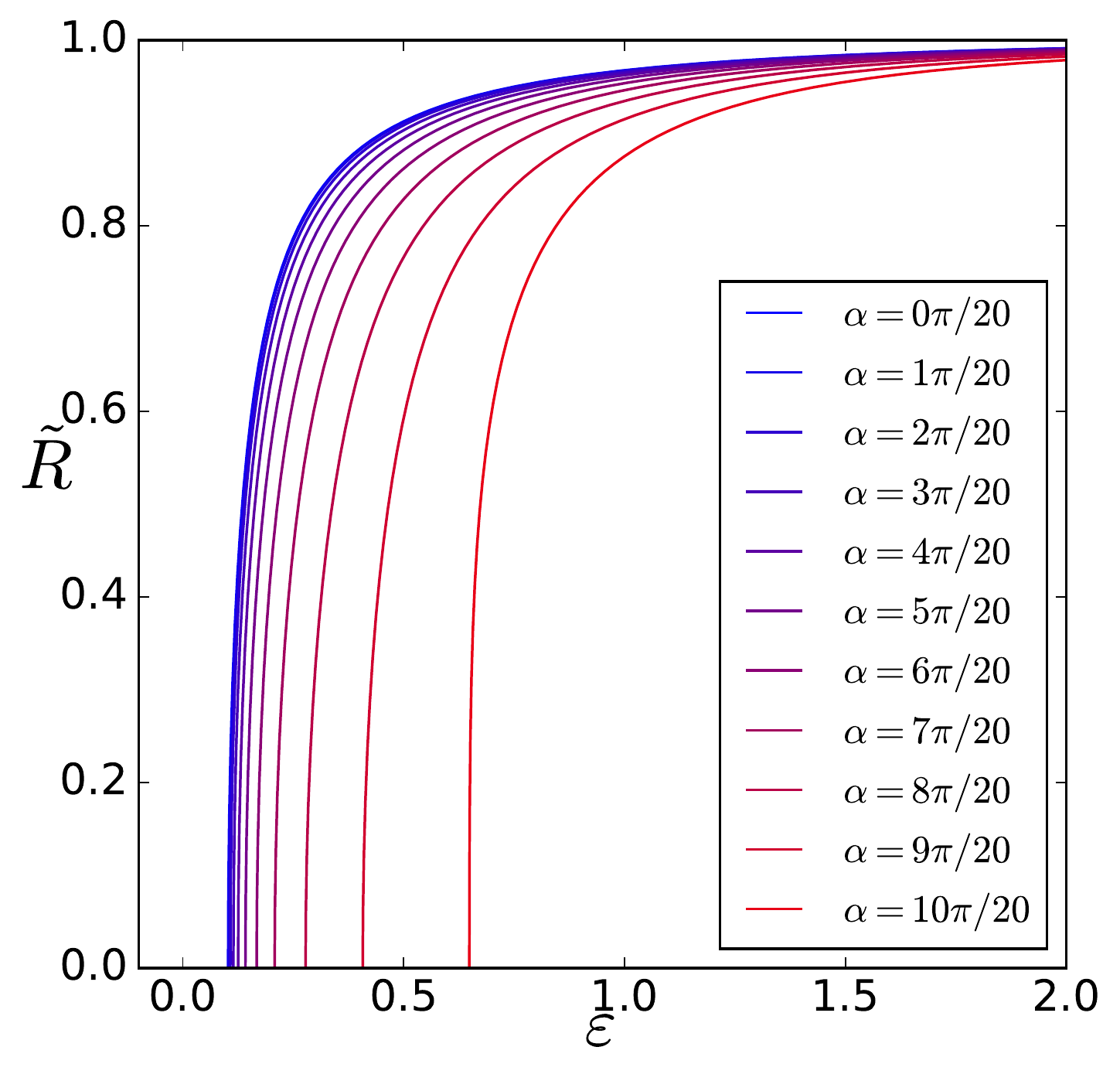}}
\put(0.1,3.8){\bf (b)}
\end{picture}
\begin{picture}(4.2,4.2)
\put(0.0,0){\includegraphics[height=4.5cm]{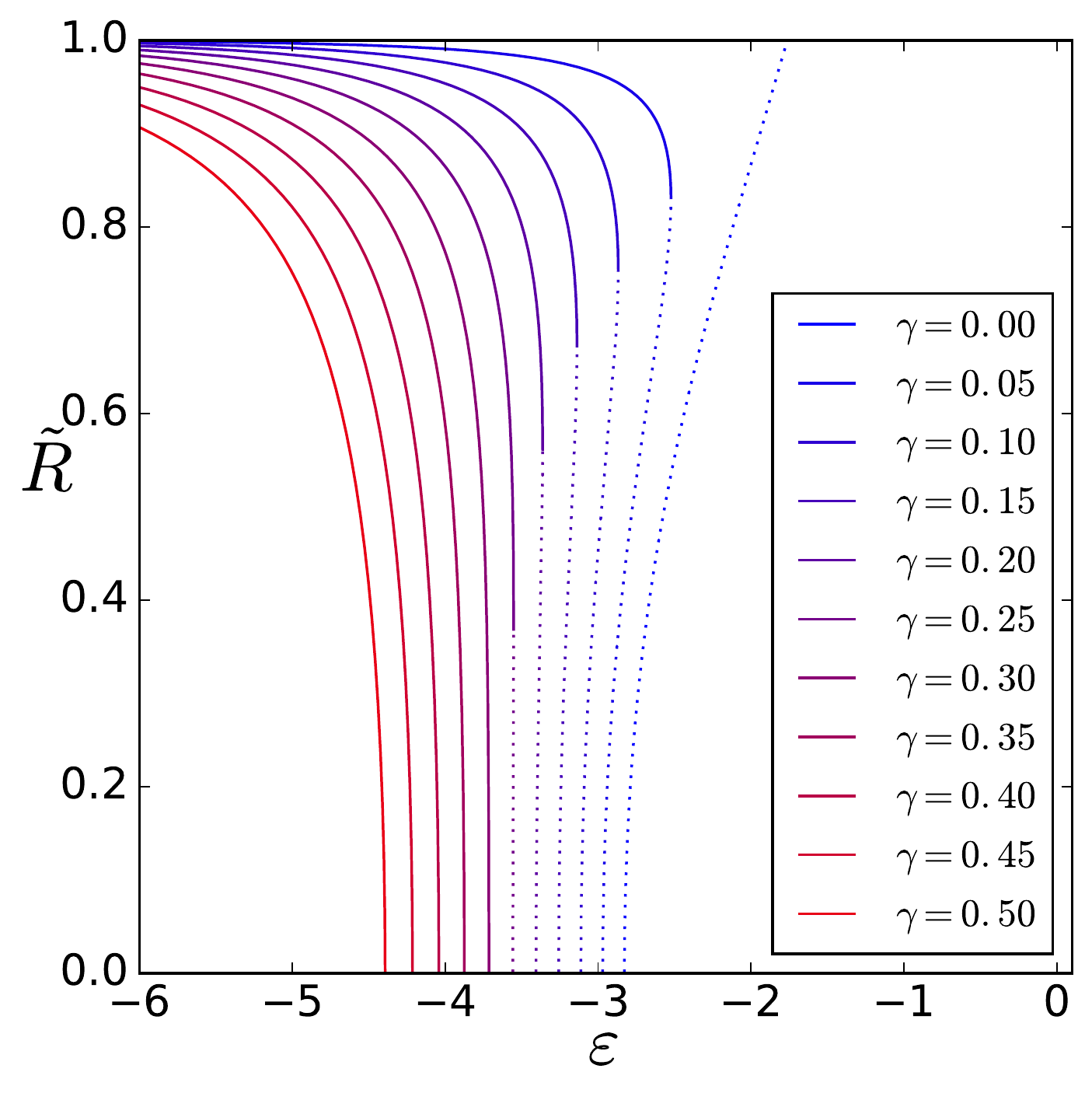}}
\put(0,3.8){\bf (c)}
\end{picture}
\begin{picture}(4.2,4.2)
\put(0.1,0){\includegraphics[height=4.4cm]{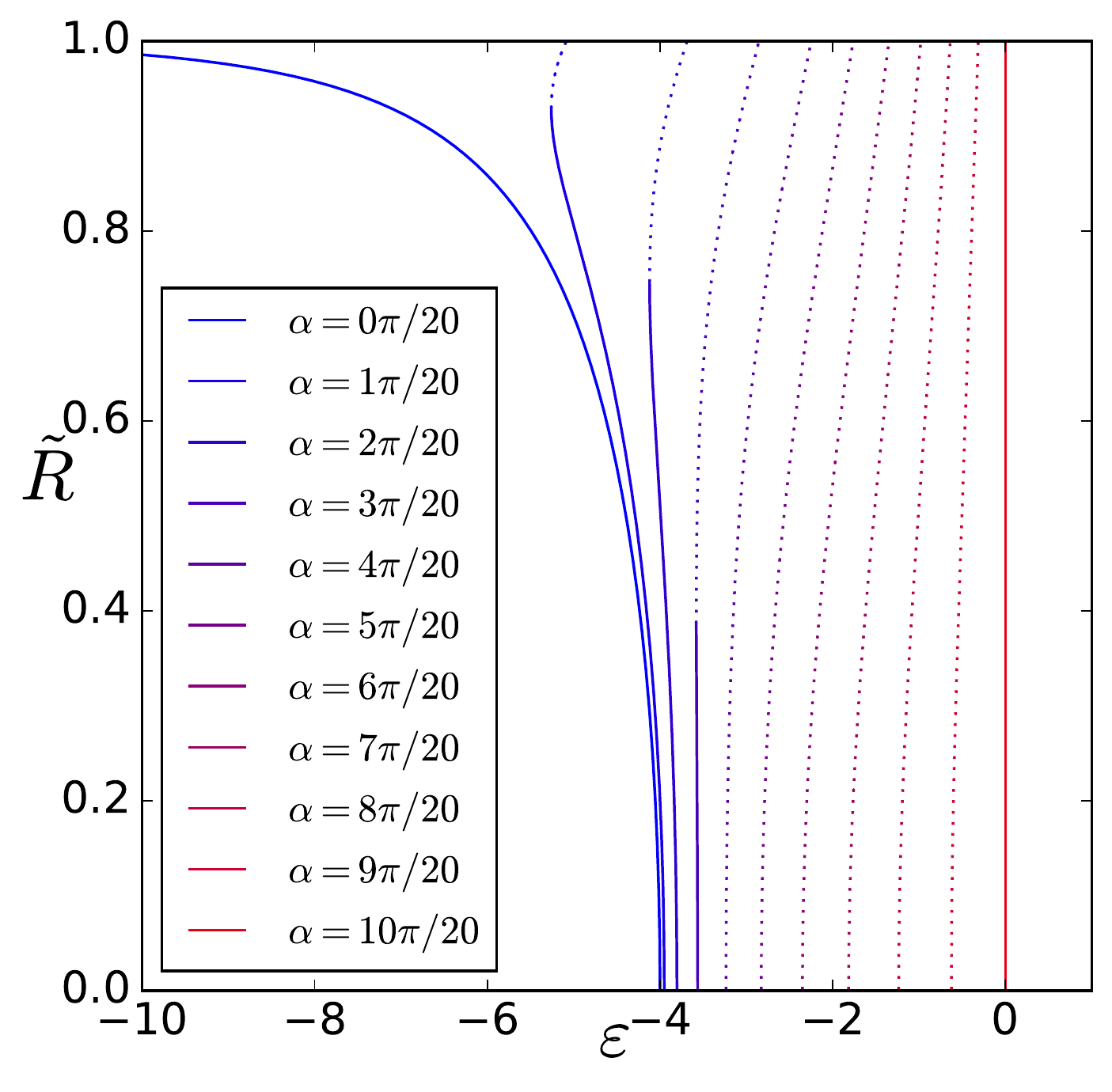}}
\put(0.1,3.8){\bf (d)}
\end{picture}
\caption{Steady state order parameter $\tilde R$ as a function of coupling strength (i.e., the bifurcation curve)
in the attractively (a)-(b) ($\varepsilon>0$) or repulsively coupled (c)-(d) ($\varepsilon<0$) M{\"o}bius map model, as shown by 
analytical expressions Eqs.~\eqref{eq:l1} and \eqref{eq:l2}. 
Without loss of generality we assume the time interval between discrete kicks to be $T = 1.0$. 
Linearly unstable and stable partially synchronized states are marked by dotted and solid lines, respectively.
In (a) and (c), we keep $\alpha=\pi/4$ constant and vary the natural frequency heterogeneity parameter $\gamma$ from zero to $0.5$ (from top to bottom). 
In (b) and (d) we set $\gamma$ to a constant value, $\gamma=0.05$ in (b) and $\gamma=0$ in (d), and vary the parameter $\alpha$. 
In (a)-(b) we see the typical second-order synchronization transition as in the classical Kuramoto-Sakaguchi model with frequency heterogeneity.
For negative coupling strengths as in (c)-(d) there can be several transitions, both continuous and discontinuous, even for identical oscillators in (d) with $\gamma=0$.
 \label{fig:bifur}} 
\end{figure}
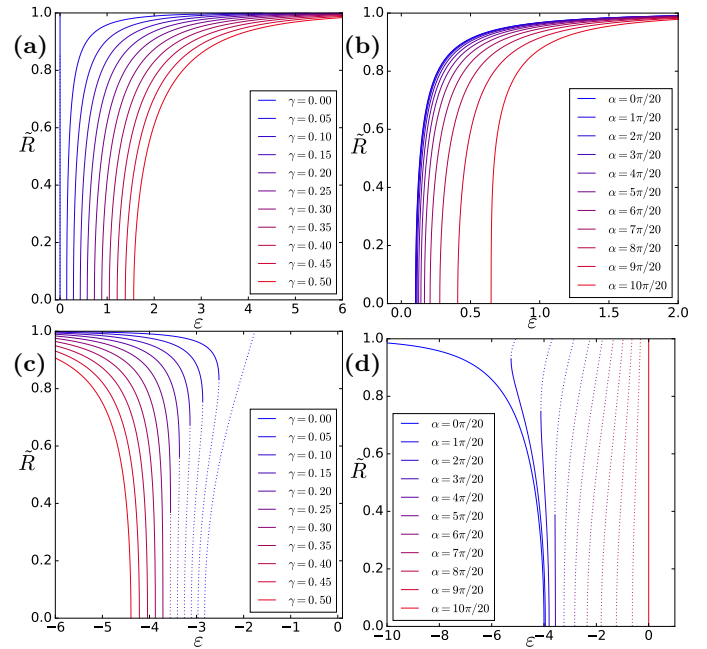

For globally coupled M{\"obius} maps we can calculate the steady state order parameter $\tilde{R} = R\exp(\gamma T)$ after each kick implicitly. Because we can always 
go into the co-rotating frame with the mean frequency $\omega_0$, we can set it to 0 without loss of generality. Setting the 
order parameter $\tilde{R}$ equal on both sides of \eqref{eq:KSMF}
\begin{equation}
\begin{gathered}
	\tilde{R}^2 = \left|\frac{\Gamma e^{-i\alpha} + \tilde{R}e^{-\gamma T}}{1 + \Gamma e^{i\alpha} \tilde{R} e^{-\gamma T}}\right|^2 = \\
	=\frac{\Gamma^2 + \tilde{R}^2e^{-2\gamma T} + 2 \Gamma \tilde{R} e^{-\gamma T} \cos\alpha}{1 + \Gamma^2\tilde{R}^2e^{-2\gamma T}+2\Gamma\tilde{R}e^{-\gamma T}\cos\alpha}\;,
	\end{gathered}
\end{equation}
where $\Gamma=\tanh\left(\varepsilon e^{-\gamma T}\tilde{R}/2\right)$, we solve a quadratic equation for $\Gamma$, and obtain
\begin{equation}
\begin{aligned}
	&\Gamma = \frac{\tilde{R}}{1-\tilde{R}^4 e^{-2\gamma  T}} \left[-(1-\tilde{R}^2) e^{-\gamma  T}\cos\alpha \right. \\
	&\left. \pm \sqrt{(1-\tilde{R}^2)^2 e^{-2\gamma  T}\cos^2\alpha + \left(1-e^{-2\gamma  T}\right)\left(1-\tilde{R}^4 e^{-2\gamma  T}\right)}\right]~.
\end{aligned}
\label{eq:l1}
\end{equation}
Inverting the expression for $\varepsilon$ we obtain
\begin{equation}
	\varepsilon = \frac{2}{\tilde{R}e^{-\gamma T}}\textrm{arctanh}\left(\Gamma\right)~.
	\label{eq:l2}
\end{equation}
Eq.~\eqref{eq:l1} and Eq.~\eqref{eq:l2} together allow us to express coupling strength $\varepsilon$ explicitly as a function of 
the steady state synchronization order parameter $\tilde{R}$ and to plot them in a bifurcation diagram, as shown in Fig.~\ref{fig:bifur}..

The first notable limit of the expression of the bifurcation curve is the existence of two critical coupling strengths for $\tilde{R}\to 0$
\begin{equation} \label{eq:bfc}
	\varepsilon_{cr} = 2\left(-\cos\alpha\pm\sqrt{\cos^2\alpha+e^{2\gamma T}-1}\right)~.
\end{equation}
Eq. \eqref{eq:bfc} implies that there is always a positive and a negative critical coupling strength for the incoherent state in globally coupled
M\"obius maps. 
The second limit is the limit of identical oscillators $\gamma\to 0$, in which case $\tilde{R}=R$ and
\begin{equation}\label{eq:bfc2}
	\Gamma = \tanh\left(\frac{\varepsilon R}{2}\right) = R \frac{-\cos\alpha \pm \left|\cos\alpha\right|}{1+R^2}~.
\end{equation}
Eq. \eqref{eq:bfc2} indicates the existence of two lines of fixed points connecting incoherence at $R=0$ and complete synchronization at $R=1$, for a given value of $\alpha$. 

Under negative coupling and identical frequency, there are several regimes for a transition to synchrony.
At $\varepsilon_0=0$, the stability of complete synchronization and incoherence is exchanged instantly. 
At
$
	\varepsilon_1 = -4\cos\alpha
$,
incoherence at $R=0$ becomes unstable, and at
$
	\varepsilon_2 = \ln\left[(1-\cos\alpha)/(1+\cos\alpha)\right]
$,
complete synchronization at $R=1$ becomes unstable. 

The existence of a synchronization transition for strongly 
repulsively coupled oscillators under discrete time stands in stark contrast to the continuous-time 
Kuramoto-Sakaguchi model \eqref{eq:kurasaka}. In the continuous case, the order parameter $R$ decreases to zero continuously under
negative coupling, whereas in the coupled-maps system a negative forcing strong enough can invert the orientation of the mean field during one step, 
and even increase its amplitude. 

Such an effect of overshooting a fixed point is typical for maps, e.g. the logistic map in contrast to the logistic differential equation. 
When interpreted physically, the map model is appropriate at describing cases where a global coupling force is implemented as sequence of pulses. 
For example, in studies of neuron populations, delayed feedback for closed-loop deep brain stimulation can induce desynchronization~\cite{Rosenblum-Pikovsky-04b,Rosenblum-Pikovsky-04c,Popovych-Hauptmann-Tass-05}. In such physiological applications, the appropriate external action on the neurons is not
of a continuous signal like in the Kuramoto model, instead it consists of a sequence of pulses~\cite{Popovych_etal_17}. The pulse is applied as a feedback through a closed loop, and the
desired amplitude of a feedback pulse is determined prior to the pulse as a function of the observed mean field. Thus, application of a pulsatile feedback~\cite{Popovych_etal_17}
can be generally described by a map of type~\eqref{eq:gcmm}. The results 
above show that a strong negative feedback may lead to a synchronization instead of desynchronization, due to the aforementioned overshooting effect.

\subsection{Two-population chimera}
\label{sec:2pop}
\begin{figure} [t]
\centering
    \includegraphics[width=.45\textwidth]{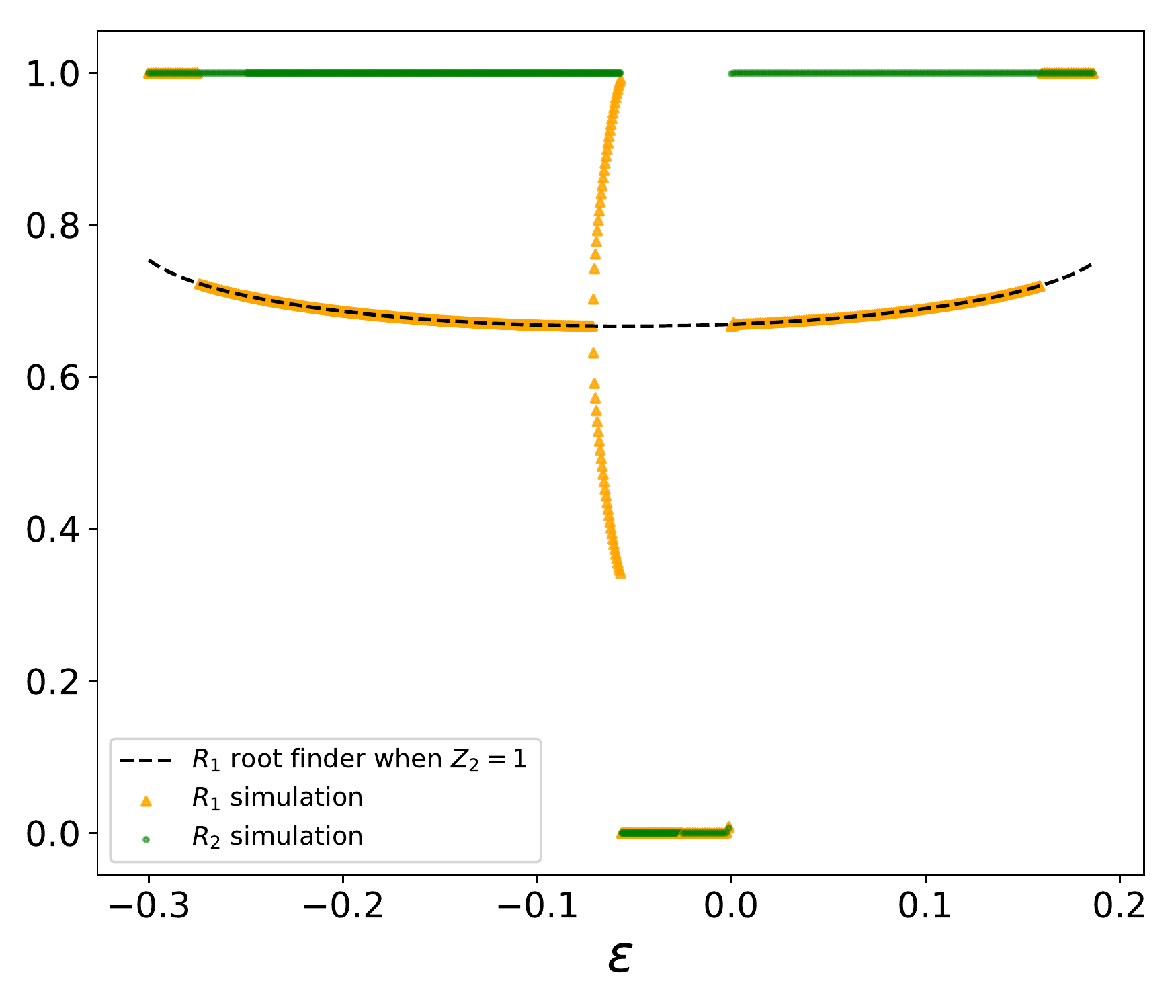}
 \caption{Bifurcation diagram illustrating the stability of the chimera states of the two 
 coupled maps of the mean fields \eqref{eq:twopopmap}. 
Scatter plots depict the stable solutions (after transient) obtained from the direct simulation
of the coupled maps \eqref{eq:twopopmap}, showing $|Z_1|$ (orange) and $|Z_2|$ (green). 
The dashed line is the fixed point of the coupled map dynamics found via numerical solver (the \texttt{findroot} function of the mpmath package~\cite{mpmath}) while
assuming one of the mean fields is 1 (at full synchrony). 
 \label{fig:bif_2pop}} 
\end{figure}
Here we consider a setup similar to the one studied in Ref.~\cite{2pop_chim}, where two populations 
of identical continuous-time oscillators
interact, with each population more strongly coupled within itself than to the other 
population. 
To formulate the corresponding M\"obius map model, we denote coupled phases in the two populations 
by their complex exponentials as before, $z_{1,j} = \exp(i\varphi_{1,j})$ and $z_{2,j} = \exp(i\varphi_{2,j})$,
and the corresponding mean field of each population as
\[
Z_1=\frac{1}{N_1}\sum_{j=1}^{N_1} z_{1,j}~,\quad Z_2=\frac{1}{N_2}\sum_{j=1}^{N_2} z_{2,j}\;.
\]
The forces acting on the populations are linear combinations of these mean fields
\begin{equation}
\begin{aligned}
   Y_1 e^{i\Psi_1}&= pZ_1 + (1-p)Z_2  \;,\\
   Y_2 e^{i\Psi_2} &= pZ_2 + (1-p)Z_1 \;,
\end{aligned}
\label{eq:dop}
\end{equation}
where parameter $p$ defines relative strengths of intra- and inter-population couplings.
The resulting M\"obius maps for the phase variables are
 \begin{equation} 
 \begin{aligned}
z^{(n+1)}_{1,j} &= \frac{z^{(n)}_{1, j}+e^{i(\Psi_1^{(n)} -\alpha)} \tanh(\frac{\varepsilon Y_1}{2})}
{1 + z^{(n)}_{1, j}e^{-i(\Psi_1^{(n)} -\alpha)} \tanh( \frac{\varepsilon Y_1}{2})} ~,\\
z^{(n+1)}_{2,j} &= \frac{z^{(n)}_{2, j}+e^{i(\Psi_2^{(n)} -\alpha)} \tanh(\frac{\varepsilon Y_2}{2})}
{1 + z^{(n)}_{2, j}e^{-i(\Psi_2^{(n)} -\alpha)} \tanh( \frac{\varepsilon Y_2}{2})} ~,
\end{aligned} \end{equation}
where $\alpha$ is the common phase shift and $\varepsilon$ is the common coupling strength. 
Here we set the identical frequency to zero by going into a co-rotating frame with the common natural frequency.

In the thermodynamical limit, 
i.e., $N_1,N_2 \rightarrow \infty$, assuming that both systems are on the OA manifold, 
we can write the dynamics of the coupled system as two 
coupled maps of the order parameters $Z_{1,2}$ (according to Eq.~\eqref{eq:oaeq})
\begin{equation} \begin{aligned} \label{eq:twopopmap}
Z^{(n+1)}_{1} &= \frac{Z_1^{(n)} + e^{i(\Psi_1 - \alpha)} \Gamma_1}{1 + Z_1^{(n)} e^{-i(\Psi_1 - \alpha)} \Gamma_1 }~,\\
Z^{(n+1)}_{2} &= \frac{Z_2^{(n)} + e^{i(\Psi_2 - \alpha)} \Gamma_2}{1 + Z_2^{(n)} e^{-i(\Psi_2 - \alpha)} \Gamma_2 }~,
\end{aligned} \end{equation}
where $\Gamma_1 = \tanh(\varepsilon Y_1 /2)$, $\Gamma_2 = \tanh( \varepsilon Y_2 /2 )$. $Y_{1,2}$ and $\Psi_{1,2}$ expressed by Eq.~\eqref{eq:dop}. For small values of coupling strength $\varepsilon$ these equations integrate the continuous-time attractively or repulsively coupled system. Effects unique to the map model can be expected for large values of $\varepsilon$. 

A bifurcation diagram of the mean-field dynamics \eqref{eq:twopopmap} is shown in Fig.~\ref{fig:bif_2pop}. 
For the numerical simulations, as in Ref.~\cite{2pop_chim}, we choose $\alpha = 0.5 \pi - 0.025$, in-group coupling ratio $p = 0.6$, 
and start iterations at initial order parameters $Z_{1,2}^{(0)}$ with small initial amplitudes, 
either close to in-phase or to anti-phase. We first evolve the coupled maps \eqref{eq:twopopmap} 
according to various positive coupling strength $\varepsilon$. At low coupling strength, depending on the initial 
conditions $Z_{1,2}^{(0)}$, we obtain either complete synchronization or chimera states, where one of the 
population is in full synchrony and the other in partial synchrony.
At high coupling strength, both populations are in globally stable full synchrony. 

For negative $\varepsilon$, we see four regimes. At $ \varepsilon_{cr}^{-}<\varepsilon$, corresponding to desynchronization by repulsive coupling in the continuous-time phase model,
we observe only the complete asynchronous case with vanishing order parameter. As we decrease $\varepsilon$ further, we see 
first a period-two chimera, then a chimera with stationary amplitudes, followed by a coexistence of 
complete synchronization and chimera, and finally complete synchronization of both populations. 
This can be contrasted again with continuous-time dynamics, where under negative coupling 
both order parameters can only decrease to zero. 

Stable amplitude chimera states are found by numeric evolution of \eqref{eq:twopopmap} and continued by root finding algorithm 
into unstable parameter regions. When we increase the negative coupling strength to values larger than $\varepsilon \approx -0.07$, a period-doubling bifurcation of the chimera 
amplitude occurs, corresponding to a periodic or quasi-periodic mean field. As $\varepsilon$ continues to increase to about $-0.06$ the quasi-periodic 
orbit collides with full synchronization and both disappear. The asynchronous state becomes stable. The loss of stability of the chimera state for a large positive coupling 
strength at $\varepsilon \approx 0.16$, similar to a large negative coupling, is again an effect of the discrete map dynamics.

\subsection{Chimera on a ring}
\label{sec:ring}
\begin{figure} [t]
\setlength{\unitlength}{1cm}
\begin{picture}(8,4.5)
\put(0,0.0){\includegraphics[height=4.5cm]{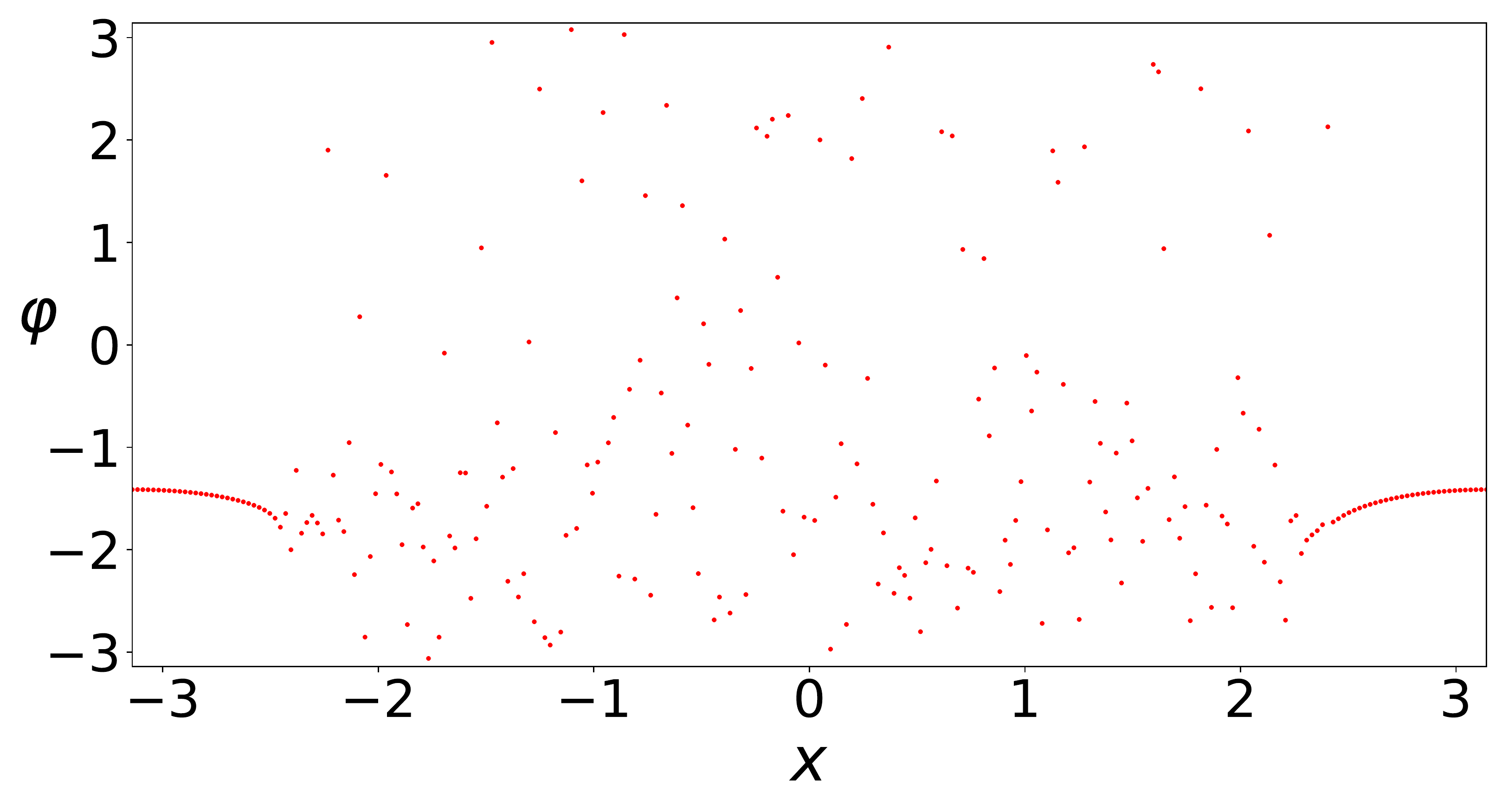}}
\put(0,3.8){\bf (a)}
\end{picture}\\
\begin{picture}(9,4.5)
\put(0,0){\includegraphics[height=4.5cm]{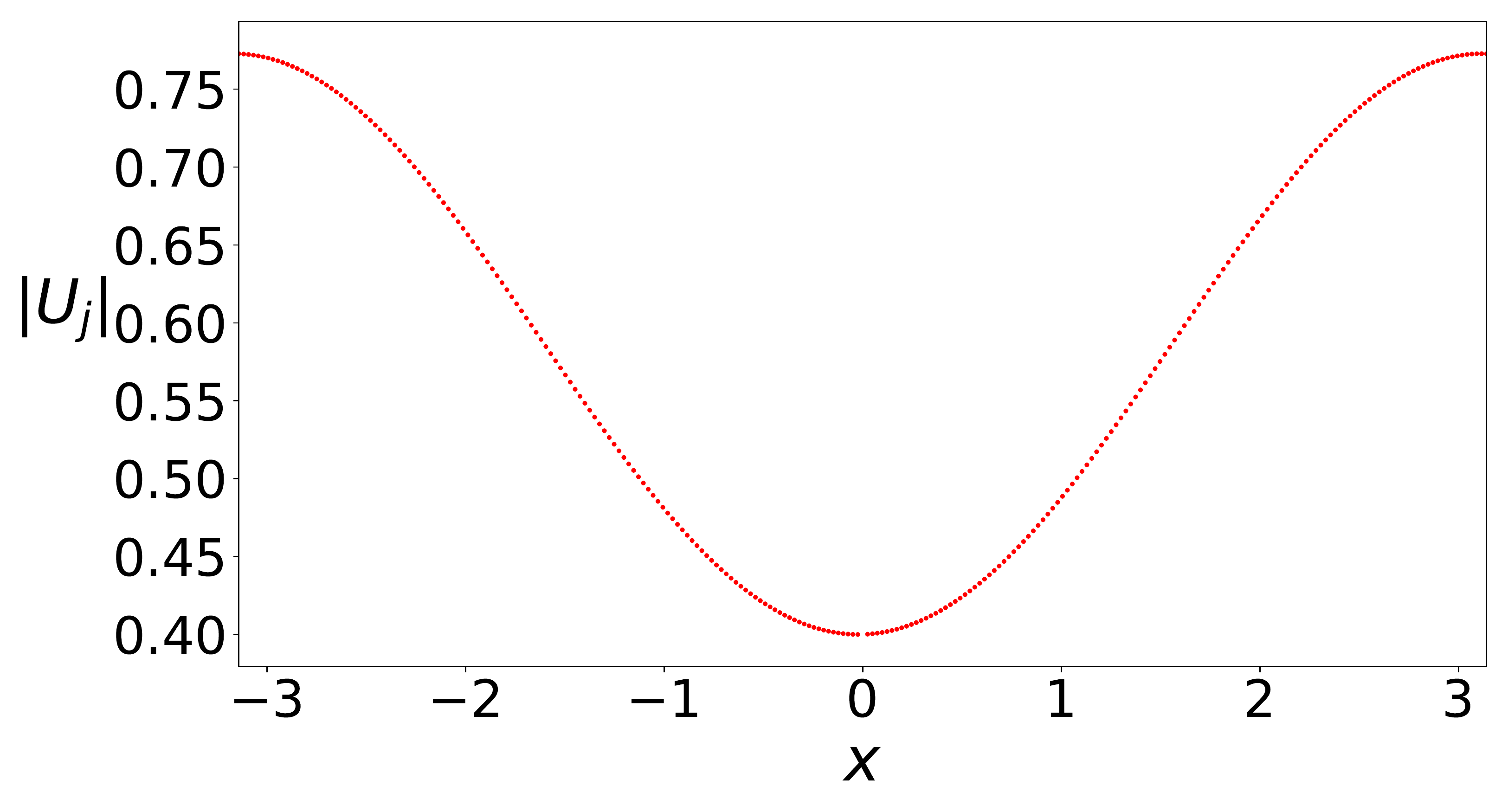}}
\put(0,3.8){\bf (b)}
\end{picture}
\caption{Chimera states on a ring in the model described by Eqs.~\eqref{eq:Zloc} and \eqref{eq:coskern}. Panel (a): 
configuration of the phases; panel (b): the local field amplitude. The chimera pattern shown here 
appears stable after 18000 steps. The network size is $N = 256$, coupling strength 
$\varepsilon = -0.8$, kernel function parameter $B = 0.995$ and phase lag $\alpha =\pi/2 - 0.18 $. 
Same as the initial condition in Ref.~\cite{ChimeraRing} for the continuous dynamics, we use 
$\varphi(t=0) = 6 r \exp(- 0.76 x^2)$, where $ r\in [-1/2, 1/2]$ is randomly sampled. 
 \label{fig:ringchim_poseps_phases}} 
\end{figure}
The first example of a chimera state for continuous-time oscillators was on a one-dimensional ring
with non-local coupling, which was explored by Kuramoto and Battogtokh~\cite{Kura_chimera,ChimeraRing}.
The oscillators are coupled via a kernel function, which determines the spatial extent
of the interactions with their neighbors. We can show similar chimera states under the coupled M{\"o}bius map model as follows.

The oscillators on the ring have positions $x_j=2\pi j /N$, where $N$ is the total number. 
Following Ref.~\cite{ChimeraRing}, we have chosen the coupling kernel as $g_{jm} = 1 + B \cos (x_j - x_m)$, so that
the complex field acting on oscillator $j$ is calculated as
\begin{equation} \begin{aligned} \label{eq:Zloc}
U_j =R_je^{i\Theta_j}= \frac{1}{N} \sum\limits^N_{m=1} g_{jm} e^{i\varphi_m} ~.\\
\end{aligned} \end{equation}
The phases are driven by these local fields according to the M{\"o}bius map
\begin{equation} \label{eq:coskern}\begin{aligned}
z^{(n+1)}_{j} = \frac{z^{(n)}_{j} + e^{i(\Theta_j^{(n)} -\alpha)} \tanh( \frac{\varepsilon R_j^{(n)}}{2})}{1 + z^{(n)}_{j}e^{-i(\Theta_j^{(n)} -\alpha)} \tanh( \frac{\varepsilon R_j^{(n)}}{2})} ~,
\end{aligned} \end{equation}
where, as before, $z_{j} = \exp (i \varphi_j)$. 

Similar to the continuous dynamics in Ref.~\cite{ChimeraRing}, we can obtain 
a stable chimera pattern for a range of positive values of $\varepsilon$ (e.g. $\varepsilon = 0.025$) 
(not shown). Same as in the two-population case before, under discrete map dynamics, there exists a regime under large
negative coupling strength which can give rise to a stable chimera pattern, see for example
Fig. \ref{fig:ringchim_poseps_phases}. 

Besides the cosine kernel function, we have also simulated the case with a square kernel, i.e., with the local field
\begin{equation} \begin{aligned} \label{eq:Zloc_sqrkern}
U_j = \frac{1}{2L+1} \sum\limits^{m=L}_{m=-L} \exp[i\varphi_{m+j}] ~.
\end{aligned} \end{equation}
Iterating map \eqref{eq:coskern} using this new local field with $N = 1000$, 
$L = 130$, $\varepsilon = 0.025$ and $\alpha = 2.71$, 
we obtained a many-headed chimera state as in the continuous case in Ref.~\cite{twisted_Chimera}. 


\section{Conclusion}
In this paper we propose a method of modelling synchronizing 
phase dynamics using a M{\"o}bius map. This map 
reproduces the dynamics of continuous-time phase oscillators.
It can be an ideal choice for fast simulation of phase synchronization, since it inherits all the properties
of continuous-time phase dynamics. In particular, neither clustering nor chaos under the iteration of a sequence of
M{\"o}bius maps can occur. All continuous-time models based on the Adler  equation, i.e. with a frequency bias and forcing proportional to the first phase harmonics, can be equivalently studied via M{\"o}bius maps. We mention here, that also phase coupling models with pure 
higher-harmonics couplings~\cite{Gong19_higherWS} can be modelled with correspondingly modified M{\"o}bius maps.

With the proposed M{\"o}bius map, we have studied map analogues of known continuous-time models for oscillator ensembles
with various connection topologies: the globally coupled Kuramoto-Sakaguchi model, two coupled populations of identical 
oscillators, and identical oscillators on a ring with non-local coupling via cosine or square distance kernel. 
For small coupling strengths and small free rotation time step, the coupled maps reproduce the dynamics of their continuous-time dynamical counterparts.
Especially, we have reproduced known chimera states with the coupled maps under non-local couplings. 
For large coupling strength, and in particular for large repulsive coupling, the discrete time dynamics
can lead to new synchronization phenomena with continuous and discontinuous bifurcations
to synchrony. This phenomenon is not observed in the equivalent continuous-time models. 


\section*{Acknowledgments}
We thank O.\;Omel'chenko for fruitful discussions. This paper is developed within the scope of the IRTG 1740/TRP 2015/50122-0, funded 
by the DFG/ FAPESP. Work of A.P. is supported by Russian Science Foundation (Grant Nr.\ 19-12-00367).

\bibliography{biblio}
\appendix
\section{M\"obius group property}
\label{sec:mgp}
The M{\"o}bius group property implies that the composition of M{\"o}bius maps is again a M\"obius map. It can be shown as follows:
\begin{eqnarray} \label{ApA}
    \mathcal{M}_{q_2,\psi_2} &=&\mathcal{M}_{q_1,\psi_1}\circ \mathcal{M}_{q_0,\psi_0}  \nonumber\\
    &=&
    \frac{q_1 + e^{i\psi_1}\frac{q_0 + e^{i\psi_0}z}{1+q_0^*e^{i\psi_0}z}}{1+q_1^*e^{i\psi_1}\frac{q_0 + e^{i\psi_0}z}{1+q_0^*e^{i\psi_0}z}} \nonumber\\
    &=& 
    \frac{q_1+q_1q_0^*e^{i\psi_0}z + q_0e^{i\psi_1} + e^{i\psi_0}e^{i\psi_1}z}{1+q_0^*e^{i\psi_0}z+q_1^*q_0e^{i\psi_1} + q_1^*e^{i\psi_1}e^{i\psi_0}z} \nonumber \\
    &=& 
    \frac{\frac{q_1+e^{i\psi_1}q_0}{1+q_1^*e^{i\psi_1}q_0}+\frac{q_1q_0^*+e^{i\psi_1}}{1+q_1^*q_0e^{i\psi_1}}e^{i\psi_0}z}{1+\frac{q_0^*+q_1^*e^{i\psi_1}}{1+q_1^*q_0e^{i\psi_1}}e^{i\psi_0}z} \nonumber\\
    &=& 
    \frac{\frac{q_1+e^{i\psi_1}q_0}{1+q_1^*e^{i\psi_1}q_0}+\frac{q_1q_0^*+e^{i\psi_1}}{1+q_1^*q_0e^{i\psi_1}}e^{i\psi_0}z}{1+\frac{q_1q_0^*+e^{i\psi_1}}{q_1q_0^*+e^{i\psi_1}}\frac{q_0^*+q_1^*e^{i\psi_1}}{1+q_1^*q_0e^{i\psi_1}}e^{i\psi_0}z} \nonumber\\
    &=& 
    \frac{\frac{q_1+e^{i\psi_1}q_0}{1+q_1^*e^{i\psi_1}q_0}+\frac{q_1q_0^*+e^{i\psi_1}}{1+q_1^*q_0e^{i\psi_1}}e^{i\psi_0}z}{1+\frac{q_1^*+e^{-i\psi_1}q_0^*}{1+q_1e^{-i\psi_1}q_0^*}\frac{q_1q_0^*+e^{i\psi_1}}{1+q_1^*q_0e^{i\psi_1}}e^{i\psi_0}z}\nonumber \\
    &=&
    \frac{\mathcal{M}_{q_1,\psi_1}(q_0) + \mathcal{C}_{q_1q_0^*}(e^{i\psi_1})e^{i\psi_0}z}{1+\mathcal{M}^*_{q_1,\psi_1}(q_0) \mathcal{C}_{q_1q_0^*}(e^{i\psi_1})e^{i\psi_0}z} \nonumber \\
    &=& \frac{q_2 + e^{i\psi_2 }z }{1+q_2^* e^{i\psi_2 }z}~.
\end{eqnarray}
A direct comparison of the last two expressions in \eqref{ApA} gives rise to the transformation rules Eqs.~\eqref{Eq:GroupParmTrans} for the group parameters
\begin{equation}
    q_2 = \mathcal{M}_{q_1,\psi_1}(q_0), \quad e^{i\psi_2}=\mathcal{C}_{q_1q_0^*}(e^{i\psi_1})e^{i\psi_0}~.
\end{equation}

\section{Dynamics of the M\"obius map}
\label{sec:app:rn}

To find the fixed points of the discrete iterated map dynamics
~\eqref{eq:mmds} with constant map parameters $q$ and $\psi$, we solve the corresponding quadratic equation
\begin{equation} \label{eq:quad}
	z^2 - \frac{e^{i\psi}-1}{q^*e^{i\psi}} z - \frac{q}{q^*e^{i\psi}} = 0 \;.
\end{equation}
Eq.~\eqref{eq:quad} has two solutions $z_1$ and $z_2$ with the properties
\begin{equation}\label{eq:quadrel}
	z_1z_2 =-\frac{q}{q^*}e^{-i\psi}\;,\quad
	z_1+z_2 = \frac{e^{i\psi}-1}{q^*e^{i\psi}}\;.
\end{equation}

From the first property it follows that $|z_1||z_2| = 1$, 
which implies that either the two fixed points are on the unit circle,
or that one fixed point is inside and the other outside the unit circle. 
According to this observation, we make the general ansatz
\begin{equation}
	z_1 = \kappa e^{i(\xi+\Delta)}\;, \qquad z_2 = \frac{1}{\kappa}e^{i\left(\xi-\Delta\right)}\;.
\end{equation}

Denoting $q=\rho \cdot \exp(i\vartheta)$ with $0 \leq \rho<1$, we obtain from \eqref{eq:quadrel} the following two relations:
\begin{eqnarray}
	\xi &=& \vartheta - \frac{\psi-\pi}{2} ~,\\
	\sin\frac{\psi}{2} &=& \frac{\rho}{2}\left[\left(\kappa+\frac{1}{\kappa}\right)\cos\Delta + i\left(\kappa-\frac{1}{\kappa}\right)\sin\Delta\right].\qquad \label{eq:2ndrelat}
\end{eqnarray}
The two fixed points do not uniquely determine the M{\"o}bius group parameters $q$ and $\psi$.
In the first regime, the two fixed points are on the unit circle, 
which implies $\kappa=1$. As a result, the second relation \eqref{eq:2ndrelat} is simplified to
\begin{equation}
	\rho\cos\Delta = \sin\frac{\psi}{2}~.
\end{equation}
The condition for fixed points on the unit circle is therefore 
\begin{equation}\label{eq:fpc2}
	\rho > \left|\sin\frac{\psi}{2}\right|~.
\end{equation}
One of the fixed points is stable and the other unstable, so under this condition the dynamics of 
the single iterated M{\"o}bius map is trivial, and the rotation number is $0$.
When equality holds in Eq.~\eqref{eq:fpc2}, it corresponds to the tangent 
bifurcation point, where the two fixed points merge into one.

In the second regime, $\kappa<1$, i.e., $z_1$ is inside the unit circle, then Eq.~\eqref{eq:2ndrelat} yields two results
\begin{eqnarray}
	\Delta &=& 0 \\
	\kappa &=& \rho^{-1}\left(\sin\frac{\psi}{2} \pm \sqrt{\sin^2\frac{\psi}{2} - \rho^{2}}\right)~.
\end{eqnarray}
For $\kappa$ to be a real number, $\rho \le |\sin(\psi/2)|$ must be satisfied, which is the
exact opposite condition from Eq.~\eqref{eq:fpc2}. Under this set of map parameters,
i.e., $\rho \le |\sin(\psi/2)|$, map ~\eqref{eq:mmds} shows rotational dynamics, which can be 
reduced to a pure rotation by virtue of a transformation which is also a M\"obius map
\begin{equation}
	y^{(n)}=\mathcal{C}_{-z_1}\left(z^{(n)}\right)
\end{equation}
The resulting pure rotational dynamics is
\begin{equation}
	y^{(n+1)}=\mathcal{C}_{-z_1}\circ\mathcal{M}_{q,\psi}\circ\mathcal{C}_{z_1}\left(y^{(n)}\right) = \mathcal{R}_{2\pi\eta}\left(y^{(n)}\right)~,
\end{equation}	
with the fixed point $z_1 = \kappa \cdot \exp(i\xi)$ as the group parameter, and the rotation number is
\begin{align} \label{eq:rot_number}
	\eta = \frac{1}{\pi}\arctan\left(\tan\frac{\psi}{2}\cdot\sqrt{1-\frac{\rho^2}{\sin^2\frac{\psi}{2}}}\right)~.
\end{align}
Eq.~\eqref{eq:rot_number} shows that in this second regime, the rotation number $\eta$ is 
a smooth function of the map parameters $\psi$ and $\rho$. 

\section{Ott-Antonsen manifold for an ensemble of M\"obius maps}
\label{sec:OAemm}
The transformation $\mu_{Q,\Psi}$ of a uniform phase density $\mu_0(\varphi)=1/(2\pi)$ under M{\"o}bius transformation $\exp(i\varphi)\to \mathcal{M}_{Q,\Psi}(\exp(i\varphi))$ of the unit circle is a wrapped Cauchy distribution
\begin{equation}\label{eq:poik}
    \mu_{Q,\Psi}(\varphi) =\mu_{Q}(\varphi) = \frac{1}{2\pi}\frac{1-|Q|^2}{|e^{i\varphi}-Q|^2}~,
\end{equation}
also known as the (univariate) Poisson kernel \cite{Marvel-Mirollo-Strogatz-09}. To show \eqref{eq:poik} is true, we can calculate the characteristic function, which are just the circular moments $\left\langle \exp(ik\varphi)\right\rangle_{\mu_{Q,\Psi}}$ of the distribution $\mu_{Q,\Psi}$, and compare that to the characteristic function of the Poisson kernel. For the circular moments of phases with density $\mu_{Q,\Psi}$, we have
\begin{eqnarray}\label{eq:mom}
	\left\langle e^{ik\varphi}\right\rangle_{\mu_{Q,\Psi}} &=& \int_0^{2\pi} e^{ik\varphi} \mu_{Q,\Psi}(\varphi)d\varphi \\
	&=& \int_0^{2\pi} \left(\mathcal{M}_{Q,\Psi}\left(e^{i\varphi}\right)\right)^k \mu_0(\varphi) d\varphi \nonumber\\
	&=& \frac{1}{2\pi}\int_0^{2\pi} \left(\frac{Q + e^{i(\Psi+\varphi)}}{1 + Q^* e^{i(\Psi+\varphi)}}\right)^k d\varphi \nonumber \\
	&=& \frac{1}{2\pi i} \oint_{|z|=1} \frac{1}{z} \left(\frac{Q + z}{1-Q^* z}\right)^k dz = Q^k~, \nonumber
\end{eqnarray}
where the last integral after the substitution $d\varphi = dz/(iz)$ is a complex contour integral with a simple pole $z=0$ inside and a $k$th-order
pole $z = (Q^*)^{-1}$ outside of the unit circle. In the derivation above we have also used the fact that the integral 
over the unit circle with respect to the transformed density $\mu_{Q,\Psi}$ is equal to the integral of the transformed circle $\mathcal{M}_{Q,\Psi}\left(S^1\right)$ with respect to the uniform density $\mu_0$.
\\ \\
From \eqref{eq:mom}, first, we see that indeed the characteristic function, and therefore the distribution $\mu_{Q,\Psi}=\mu_Q$ is independent of $\Psi$. Second, the circular moments are integer powers of the M\"obius map parameter $Q$, and in particular, the first moment $Z=\left\langle \exp(i\varphi)\right\rangle_{\mu_Q} = Q$. 
\\ \\
The circular moments of the Poisson kernel can be similarly obtained via complex integration. Given $z=\exp(i\varphi)$ and $d\varphi=dz/(iz)$, it follows
\begin{eqnarray}
    \left\langle e^{ik\varphi}\right\rangle &=& \frac{1}{2\pi} \int_0^{2\pi}  e^{ik\varphi}\frac{1-|Q|^2}{|e^{i\varphi}-Q|^2}d\varphi\\ 
    &=& \frac{1}{2\pi} \int_0^{2\pi} \frac{e^{i(k+1)\varphi}(1-|Q|^2)}{(e^{i\varphi}-Q)(1-Q^*e^{i\varphi})}d\varphi \nonumber \\
    &=& \frac{1}{2\pi i} \oint_{|z|=1} \frac{z^k (1-|Q|^2)}{1-Q^*z} \frac{1}{(z-Q)} dz = Q^k ~. \nonumber
\end{eqnarray}
\\ \\
Since they have the same characteristic function, the density function $\mu_Q$ must be identical to the Poisson kernel, i.e. Eq. \ref{eq:poik}.
\section{Solution of Adler equation over a finite time interval}
\label{sec:km}
Here we will show that the kick map 
\begin{equation} \label{eq:kickmap}
	\mathcal{K}_{\lambda,A\tau,\beta}\left(z\right) = \frac{(\sigma + i\lambda\Gamma)z + e^{i\beta}\Gamma}{(\sigma - i\lambda\Gamma) + z e^{-i\beta}\Gamma}
\end{equation}
with $\sigma = \sqrt{1-\lambda^2}$ and $\Gamma = \tanh(A\tau\sigma/2)$ under the conventions $\sqrt{-1}=i$ and $\tanh(ix)=i\tan(x)$ is indeed a solution of the Adler equation
\begin{equation} \label{ap:eqAd}
    \dot\varphi = A\left[\lambda - \sin(\varphi-\beta)\right]
\end{equation}
with constant parameters $A$, $\lambda$ and $\beta$. With $z=e^{i\varphi}$, \eqref{ap:eqAd} can be written in a complex form
\begin{equation}\label{eq:complex_phase}
    \dot z = iz A\left(\frac{1}{2i}e^{i\beta}z^* + \lambda -\frac{1}{2i}e^{-i\beta}z\right) ~.
\end{equation}
The parameters of the kick map can be directly taken from the Adler equation, and the kick map can be easily cast into the canonical form \eqref{eq:mm} of the M{\"o}bius map, where
\begin{equation}
	q= e^{i\beta}\frac{\Gamma}{\sigma-i\lambda\Gamma}~, \qquad e^{i\psi} = \frac{\sigma+i\lambda\Gamma}{\sigma-i\lambda\Gamma}~. 
\end{equation}
Note that $\sigma$ and $\Gamma$ are either both real or both imaginary. The complex conjugates of $q$ and $\exp(i\psi)$ are therefore obtained by just replacing $i\to -i$ in the formulas above.
Without loss of generality we can make a substitution $z\to z\exp(i\beta)$, $q\to q\exp(i\beta)$  and consider only $\beta=0$. Marvel et al. \cite{Marvel-Mirollo-Strogatz-09} have shown that for phase equations \eqref{eq:complex_phase}, with arbitrary time dependence of the parameters, the solution is given by a M{\"o}bius transform where the parameters evolve according to the ordinary differential equations
\begin{eqnarray}
    \dot q &=& A\left(\frac{1}{2}+i\lambda q - \frac{1}{2}q^2\right) \label{ap:eq1} \\
    \dot \psi &=& A\left(\frac{1}{2i}q^*+\lambda  - \frac{1}{2i}q\right) \label{ap:eq2}~.
\end{eqnarray}
Thus, we just need to show that the same is true for the parameters $q$ and $\psi$ and for our choice of $\Gamma(\tau)$. 

The left-hand-side of \eqref{ap:eq1} can be calculated as
\begin{equation}
    \dot q = \frac{\sigma \dot\Gamma}{(\sigma-i\lambda\Gamma)^2}
    = \frac{\sigma^2 \frac{A}{2}(1-\Gamma^2)}{(\sigma-i\lambda\Gamma)^2} ~,
\end{equation}
which matches the right-hand-side
\begin{eqnarray}
    &&A\left(\frac{1}{2} + i\lambda q - \frac{1}{2}q^2 \right) \\
    &=& \frac{A}{(\sigma -i\lambda\Gamma)^2}
    \left(\frac{1}{2}(\sigma -i\lambda\Gamma)^2 + i\lambda\Gamma(\sigma -i\lambda\Gamma) - \frac{1}{2}\Gamma^2\right) \nonumber \\
    &=& \frac{A}{(\sigma -i\lambda\Gamma)^2}
    \left(\frac{1}{2}\sigma^2+\frac{1}{2}\lambda^2\Gamma^2-\frac{1}{2}\Gamma^2\right) \nonumber \\
    &=& \frac{\frac{A}{2}}{(\sigma -i\lambda\Gamma)^2}\left(\sigma^2-(1-\lambda^2)\Gamma^2\right) = \frac{\sigma^2 \frac{A}{2}(1-\Gamma^2)}{(\sigma-i\lambda\Gamma)^2} ~.\nonumber 
\end{eqnarray}
This proves the first identity \eqref{ap:eq1}. Similarly, we observe 
\begin{equation}
    \frac{d}{d \tau} e^{i\psi} = \frac{2i\lambda\sigma\dot\Gamma}{(\sigma-i\lambda\Gamma^2)} 
    = i\lambda A e^{i\psi}\frac{\sigma^2(1-\Gamma^2)}{\sigma^2+\lambda^2\Gamma^2}
\end{equation}
and
\begin{eqnarray}
    && ie^{i\psi}A\left(\frac{1}{2i}q^*+\lambda-\frac{1}{2i}q\right) \\
    &=& iA e^{i\psi}\left(\frac{1}{2i}\frac{\Gamma}{\sigma+i\lambda\Gamma}+\lambda-\frac{1}{2i}\frac{\Gamma}{\sigma-i\lambda\Gamma}\right) \nonumber \\
    &=&
    iA e^{i\psi}\frac{\frac{1}{2}\Gamma(\sigma-i\lambda\Gamma)+i\lambda(\sigma^2+\lambda^2\Gamma^2)-\frac{1}{2}\Gamma(\sigma+i\lambda\Gamma)}{i(\sigma^2+\lambda^2\Gamma^2)} \nonumber \\
    &=&
    i A e^{i\psi}\frac{\lambda(\sigma^2+\lambda^2\Gamma^2)-\lambda\Gamma^2}{\sigma^2+\lambda^2\Gamma^2}
    = i\lambda A e^{i\psi}\frac{\sigma^2(1-\Gamma^2)}{\sigma^2+\lambda^2\Gamma^2}  ~,\nonumber 
\end{eqnarray}
which proves the second identity \eqref{ap:eq2}.
\\ \\
Unlike those of the standard parametrization of the M\"obius map \eqref{eq:mm} (Appendix \eqref{sec:app:rn}), the fixed points of the kick map \eqref{eq:kickmap} have a more direct relation to the parameters. The fixed point equation
\begin{equation}
    z = \frac{(\sigma+i\lambda\Gamma)z + \Gamma}{(\sigma-i\lambda\Gamma)+z\Gamma}
\end{equation}
can be changed into the quadratic form $ z^2 - 1 - 2i\lambda z = 0$ which is solved by
\begin{equation}
    z = i\lambda \pm \sqrt{1-\lambda^2} = i\lambda \pm \sigma~.
\end{equation}
If $|\lambda|<1$, the fixed points of the kick map are located on the unit circle at phases $\varphi$ with $\sin\varphi=\lambda$ and $\cos\varphi=\pm\sigma$. If $|\lambda|>1$, the fixed points are on the imaginary axis inside and outside the unit circle at $i(\lambda\pm\sqrt{\lambda^2-1})$. In both cases the locations of the fixed points are independent of $\Gamma$, i.e. of the kick map parameter $A\tau$, which only controls the degree of contraction, expansion or rotation around the fixed points and not their locations.
\section{Computational efficiency of M\"obius map models}
\label{sec:ce}
For a single iterated map, due to the complexity inherent in the algebraic form of the map, a single step with the M{\"o}bius map Eq.~\eqref{eq:atan2} is 2.5 times slower than the sine map, or equivalently, than an Euler integration step of the corresponding ODE system \eqref{eq:adleq}. However, when integrating Adler equation, whose property is inherited by the equivalent M{\"o}bius map, the time step $\tau$ can be large and a significant speed-up can be achieved.

Nevertheless, for systems of globally or locally coupled oscillators, the main bottleneck in terms of computational efficiency is the calculation of the mean fields. This bottleneck remains also under the application of M\"obius maps. Euler integration and M\"obius map each requires one calculation of the mean field in every step which takes about the same time. This is confirmed by double precision Euler integration of the Kuramoto-Sakaguchi Model Eq.~\eqref{eq:kurasaka}, as compared to simulations with M\"obius map model Eq.~\eqref{eq:gcmm}. When Runge-Kutta integration is used, which requires four evaluations of the mean field per integration step, the method using M\"obius map is four times faster, similar to Euler scheme. 

However, while Euler scheme violates partial integrability of the original globally coupled ODE dynamics, M\"obius map preserves such property \cite{Gong19}. Certainly, many synchronization effects can also be observed in the integrations of ODEs with large integration steps, even with correspondingly large integration errors. The main distinction of M\"obius map from the integration of ODEs is the invariance of the OA manifold. When a more precise solution of the ODE system is required, the time step usually needs to decrease by a factor of 10 or 100, which makes integrating ODE correspondingly much slower compared to evolving the system using M\"obius maps. 

In fact, using Watanabe-Strogatz theory, we can combine M{\"o}bius maps with a numerical integration of the reduced quasi-mean-field equations, i.e. ODEs \eqref{ap:eq1} and \eqref{ap:eq2}, for the M{\"o}bius group parameters to obtain ODE solutions of the full system to desired precision that conserve all constants of motion (see Ref.~\cite{Marvel-Mirollo-Strogatz-09} and a practical example in Ref.~\cite{Gong19_higherWS}) but the necessary calculation of the mean field from the constants of motion is computationally more involved than an integration with regular numerical schemes, because an additional transformation from the constants to the phases is needed at every integration step.
\end{document}